\documentclass[aps,prd,showpacs,preprintnumbers,groupedaddress]{revtex4}
\usepackage{amsmath}
\usepackage{graphicx}
\usepackage{bm}
\usepackage{color,colordvi}
\begin{document}

\newcommand{\as}{\alpha_{\textrm s}}

\def\plus{{\!+\!}}
\def\minus{{\!-\!}}
\def\z#1{{\zeta_{#1}}}
\def\zz#1{{\zeta_{#1}^2}}

\def\ca{{C^{}_A}}
\def\cf{{C^{}_F}}
\def\tr{{T^{}_{\! R}}}

\def\caca{{C^{2}_A}}
\def\cfcf{{C^{2}_F}}

\def\nf{{n^{}_{\! f}}}
\def\n2f{{n^{\,2}_{\! f}}}

\def\nc{{N^{}_{\! c}}}
\def\n2c{{N^{\,2}_{\! c}}}

\def\li#1{{\textrm{Li}_{#1}}}
\def\h{{\textrm{H}}}
\def\D#1{{\mathcal{D}_{#1}(z)}}

\preprint{YITP-SB-04-27}
\preprint{BNL-HET-04/4}
\preprint{hep-ph/0405219}
\title{Next-to-leading Log Resummation of Scalar and Pseudoscalar Higgs \\
       Boson Differential Cross-Sections at the LHC and Tevatron}
\author{B. Field}
\email[]{bfield@ic.sunysb.edu}
\affiliation{C.N. Yang Institute for Theoretical Physics, 
             Stony Brook University,
             Stony Brook, New York 11794-3840, USA}
\affiliation{Department of Physics, Brookhaven National Laboratory,
             Upton, New York 11973, USA}
\date{May 22, 2004}

\begin{abstract}

The region of small transverse momentum in $q\bar{q}-$ and
$gg-$initiated processes must be studied in the framework of resummation
to account for the large, logarithmically-enhanced contributions to
physical observables. In this paper, we will calculate the fixed order
next-to-leading order (NLO) perturbative total and differential
cross-sections for both a Standard Model (SM) scalar Higgs boson and the
Minimal Supersymmetric Standard Model's (MSSM) pseudoscalar Higgs boson
in the Heavy Quark Effective Theory (HQET) where the mass of the top
quark is taken to be infinite. Resummation coefficients $B^{(2)}_g,
C^{(2)}_{gg}$ for the total cross-section resummation for the
pseudoscalar case are given, as well as $\bar{C}^{(1)}_{gg}$ for the
differential cross-section.

\end{abstract}

\maketitle

\section{Introduction} 
\label{intro}

The discovery of one or more Higgs bosons is the central research
interest for high energy physics programs at hadron colliders around the
world. Beyond the phenomenology of the Standard Model (SM) Higgs boson,
the Minimal Supersymmetric Standard Model (MSSM) which is a special case
of the Two Higgs Doublet Model (2HDM) is of particular interest to
theorists. For a review see Ref.~\cite{Gunion:1989we, Carena:2002es}.

Very recently, a new central value for the top quark mass was
reported\cite{Group:2004rc}. This changed the exclusion limits on the SM
Higgs boson, putting its central mass value from precision electroweak
fits at $117$~GeV/c$^2$. This is exciting because this value is above
the exclusion limits from the LEP direct searches which exclude the mass
of a SM-like Higgs boson below approximately
$114$~GeV/c$^2$\cite{Barate:2003sz}. In the MSSM there are five physical
Higgs bosons; a light and a heavy scalar ($h^0$,$H^0$), two charged
scalars ($H^\pm$), and a CP-odd pseudoscalar ($A^0$). The mass of the
pseudoscalar is excluded\cite{Fernandez:2003hu} from being lighter than
$92$~GeV/c$^2$. The ratio between the vacuum expectation values
(\textsc{vev}s) of the two neutral Higgs bosons of the MSSM is defined
as $\tan\beta = v_2/v_1$. For $m_{\textrm{top}} = 174.3$~GeV/c$^2$, $0.5
< \tan\beta < 2.4$ has been excluded by the LEP Higgs searches. The
bounds on $\tan\beta$ will change as the central value of the top mass
changes. A full analysis of the $\tan\beta$ exclusion bounds for the new
top mass of $m_{\textrm{top}} = 178.0$~GeV/c$^2$ is not yet available.
In this paper, we will leave $\tan\beta = 1$ so that the pseudoscalar
results can easily be scaled by an appropriate number of interest to the
reader, and the mass of the Higgs bosons will be set to $M_\Phi =
120$~GeV/c$^2$, where $\Phi$ is the Higgs boson of interest.

In the context of resummation, the literature has focused on the scalar
Higgs boson\cite{Catani:ne, Kauffman:1991jt, Yuan:1991we, Kauffman:cx,
Kramer:1996iq, Catani:1996yz, Balazs:2000wv, deFlorian:2000pr,
deFlorian:2001zd, Glosser:2002gm, Berger:2002ut, Berger:2003pd,
Bozzi:2003jy, Catani:2003zt, Kulesza:2003wn}. This paper will provide
resummation coefficients for the pseudoscalar Higgs boson for the total
cross-section and differential distributions. Our calculations are done
in the Heavy Quark Effective Theory (HQET) where the mass of the top
quark is taken to be infinite. The role of the bottom quark in
pseudoscalar production becomes dominant at large $\tan\beta$. In order
to correctly take the bottom quark into account at this order, massive
resummation coefficients will have to be determined. This will be
reserved for another discussion. 

In Section~\ref{HQET}, we will introduce the Heavy Quark Effective
Theory (HQET) in which our calculations were performed. In
Section~\ref{resum}, we will introduce our resummation conventions and
present our new results for pseudoscalar resummation. Finally, in
Section~\ref{results} we will present our numeric results for the
differential distributions for the SM scalar and MSSM pseudoscalar Higgs
boson at the Large Hadron Collider (LHC) and Tevatron in the HQET.

\section{Heavy Quark Effective Theory}
\label{HQET}

Higgs phenomenology in QCD lends itself well to the use of Heavy Quark
Effective Theory (HQET). When only the top quark is considered in
calculations, it is possible to replace the top quark loops by an
effective vertex when the other quarks are ignored. The role of the
bottom quark in Higgs physics has recently been examined in great
detail\cite{Harlander:2003ai, Field:2003yy}, but will not be included in
this paper. The Lagrangian that describes this effective vertex can be
derived from the $gg \rightarrow \Phi$ (where $\Phi$ is a Higgs boson of
interest) triangle diagram\cite{Wilczek:1977zn, Ellis:1979jy,
Georgi:1977gs, Rizzo:1979mf} and letting the mass of the quark become
infinitely heavy at the end of the calculation\cite{Shifman:eb,
Vainshtein:ea, Voloshin:hp}. The Higgs of interest could be the SM
scalar Higgs, or the pseudoscalar Higgs of the MSSM. In principle, the
light and heavy scalar Higgs of the MSSM can be included in this
formalism in place of the SM Higgs by multiplying by the appropriate
coupling factor. However, supersymmetric corrections within the MSSM
will not be included in this paper, only SM QCD corrections will be
included.

The HQET method allows for the inclusion of the $\mathcal{O} (\epsilon)$
terms which are important for deriving resummation coefficients. This
program leads to the effective Lagrangian in $d=4-2\epsilon$ dimensions
for a scalar Higgs boson

\begin{equation}
\mathcal{L}_{\textrm{eff}}^H
            = -\frac{1}{4} g_H H G^{a}_{\mu \nu} G^{a,\mu \nu}
             \biggl( \frac{4\pi\mu_r^2}{m_\textrm{top}^2}\biggr)^\epsilon
             \Gamma(1+\epsilon) 
\end{equation}
where $g_H = \as / 3\pi v$ is the coupling of the effective vertex at
LO, $G^{a,\mu\nu}$ is the field-strength tensor for the gluons, $\mu_r$
is the renormalization scale, and the vacuum expectation value
(\textsc{vev}) of the Higgs is defined as $v^2 = (\sqrt{2}G_F)^{-1}
\simeq 246$~GeV. The effective coupling receives order by order
corrections that have been calculated\cite{Chetyrkin:1997sg,
Kramer:1996iq} previously. The appearance of the top quark mass in this
expression hints that the corrections to the effective coupling at
higher order may include logarithmic corrections including the top quark
mass, which is the case. Alternatively, we can define $1/v^2 =
6422.91$~pb, which is convenient in cross-section calculations. This
effective Lagrangian generates effective vertices with two, three, and
four gluons with a scalar Higgs boson. The Feynman rules for a scalar
Higgs can be found in the literature\cite{Kauffman:1996ix}.

When a pseudoscalar Higgs boson is considered, the effective LO
Lagrangian changes due to the $\gamma_5$ coupling and can be written
\begin{equation}
\mathcal{L}_{\textrm{eff}}^{A^0}
            =\frac{1}{4} g_{A^0} A^0 G^{a}_{\mu \nu} 
             \tilde{G}^{a,\mu \nu}
             \biggl( \frac{4\pi\mu_r^2}{m_\textrm{top}^2}\biggr)^\epsilon
             \Gamma(1+\epsilon)
\end{equation}
where $g_{A^0} = \as / 2\pi v$ is the coupling of the effective vertex.
The $\tilde{G}^{a,\mu \nu} = 1/2 \, \epsilon^{\mu\nu\rho\sigma}
G_{\rho\sigma}^a$ operator is the dual of the usual gluon field-strength
tensor. It is important to note that the four gluon plus pseudoscalar
Higgs vertex is absent in this effective Lagrangian as its Feynman rule
is proportional to a completely antisymmetric combination of structure
functions and therefore vanishes. It should be noted that this is the LO
effective Lagrangian, and that a second operator begins to contribute at
higher orders. A complete discussion can be found in
Ref.~\cite{Chetyrkin:1998mw}. The Feynman rules for the pseudoscalar can
be found in the literature\cite{Kauffman:1993nv}. The Feynman rules
listed in this reference should be used with $g_{A^0} = \as / 2\pi v$ to
avoid a spurious extra factor of $1/4$.

If we define $z=M_\Phi^2/\hat{s}$ where $\hat{s}$ is the partonic center
of momentum energy squared, the partonic total cross-section for Higgs
production (either scalar or pseudoscalar written here generically as
$\Phi$) from the fusion of two partons $a$ and $b$ can be written as a
series expansion in $\as$ as

\begin{align} \nonumber
    \hat{\sigma}(ab \rightarrow \Phi + X)
& = \hat{\sigma}_0^\Phi \Delta_{ab \rightarrow \Phi} \\
& = \hat{\sigma}_0^\Phi
\biggl( 
        \delta_{ag}\delta_{bg}\delta(1-z) +
        \frac{\as}{\pi} 
        \Delta^{(1)}_{ab \rightarrow \Phi}(z) +
        \biggl( \frac{\as}{\pi} \biggr)^2 
        \Delta^{(2)}_{ab \rightarrow \Phi}(z) +
        \mathcal{O}(\as^3) + \cdots
\biggr),
\end{align}
where $\hat{\sigma}_0^\Phi$ is the LO partonic total cross-section, and
the $\Delta^{(n)}_{ab \rightarrow \Phi}(z)$ coefficients are higher
order corrections.

The NNLO corrections for both the scalar and pseudoscalar Higgs bosons
have been calculated in the HQET\cite{Harlander:2001is,
Harlander:2002vv, Ravindran:2003um}. Although the next-to-leading order
corrections have been calculated in several places\cite{Dawson:1990zj,
Djouadi:1991tk, Kauffman:1991jt, Kauffman:cx, Kauffman:1993nv,
Spira:1995rr, Ravindran:2002dc, Field:2002pb}, there are some
discrepancies in the literature that we would like to clear up in this
paper. For this reason, we will explicitly calculate the NLO corrections
for the $gg$ initial state. The importance of this particular channel,
beyond its relevance to high-energy hadron collisions, will be addressed
later. The partonic cross-section at NLO has to be written as the sum of
the real emissions, the virtual corrections, the charge renormalization,
and Altarelli-Parisi subtractions as follows

\begin{equation}
\hat{\sigma}_{\text{NLO}} = \hat{\sigma}_{\text{real}}
                          + \hat{\sigma}_{\text{virt}}
                          + \hat{\sigma}_{\text{ren}}
                          + \hat{\sigma}_{\text{AP}}.
\end{equation}

Although a few very thorough treatments for the scalar case exist in the
literature\cite{Dawson:1990zj, Djouadi:1991tk}, we will re-derive them
here to highlight differences in the pseudoscalar case, where an
exhaustive treatment is missing. We will follow closely the discussion
in these references. We will see that only the $gg$ channel will play a
role in the resummation formalism.

\subsection{Scalar Higgs Matrix Elements}
\label{scalarME}

Suppressing $\mathcal{O}(\epsilon)$ terms for now, the matrix elements
for scalar Higgs production in the HQET can be written
as\cite{Dawson:1990zj, Djouadi:1991tk, Kauffman:1991jt, Kauffman:cx,
Kauffman:1993nv, Spira:1995rr, Ravindran:2002dc, Field:2002pb}

\begin{equation}
\mathcal{M}(g(p_1^{\mu,A})+g(p_2^{\nu,B})\rightarrow H)
 = - \frac{\as}{3\pi v}\delta^{AB} \biggl( \eta^{\mu\nu} 
     \frac{M_H^2}{2} - p_1^\nu p_2^\mu \biggr) \epsilon_\mu(p_1)
     \epsilon_\nu(p_2).
\end{equation}
Here $AB(\mu\nu)$ are the color (Lorentz) indices of the incoming
gluons. When the matrix elements are squared, the contraction of the
delta functions yields a factor of $\delta^{AB} \delta_{AB} = N_c^2 - 1
= 8$, the contraction of the metric yields $\eta^{\mu\nu}\eta_{\mu\nu} =
d$. This yields the squared matrix elements (before any color-spin
averaging)

\begin{equation}
|\mathcal{M}(gg\rightarrow H)|^2 
               = \frac{\as^2 M_H^4 (N_c^2-1)}
                      {16 \pi^2 v^2}
                  \biggl( \frac{4\pi\mu_r^2}{m_\textrm{top}^2} 
                  \biggr)^{2\epsilon}
                  \Gamma^2(1+\epsilon)(1-\epsilon).
\end{equation}

In this $2 \rightarrow 1$ process, it is easiest to calculate the decay
width of the Higgs and convert that to a partonic total cross-section. 
We need to color and spin average the matrix elements squared noting
that the gluon-gluon initial state must be averaged over
$4(1-\epsilon)^2$ transverse polarizations. We can write the partonic
cross-section in terms of the decay width and simply add $\delta(1-z)$
since $z = M_H^2/\hat{s} = 1$ at threshold, so that\cite{Dawson:1990zj}

\begin{align}
\hat{\sigma}_0^H 
               (gg\rightarrow H) & = \frac{\pi^2}{8 M_H^3} \,
                                  \Gamma(H \rightarrow gg) \,
                                  \\ \nonumber
            & = \biggl( \frac{\as}{\pi} \biggr)^2
                \frac{\pi}{576 v^2}
                \biggl( \frac{4\pi\mu_r^2}{m_\textrm{top}^2} 
                \biggr)^{2\epsilon}
                \frac{\Gamma^2(1+\epsilon)}
                     {1-\epsilon}.
\end{align}
This factor (with its $\epsilon$ dependence) will be pulled from each of 
the higher order correction factors. The LO cross-section in the 
HQET starts at $\mathcal{O}(\as^2)$.

\subsubsection{Radiative Corrections}
\label{scalarRad}

At the next order in perturbation theory, there are $qg$ and $q\bar{q}$ 
initial state processes. However, as we will see later, we are only 
interested in the $gg$ initial state, so we will only calculate the NLO 
corrections to the $gg$ initial state process.

The real contributions at NLO to the $gg$ initial state come from the
process $gg \rightarrow gH$ which was originally calculated in the
$\epsilon \rightarrow 0$ limit in the full theory including a finite
mass top quark in Ref.~\cite{Ellis:1987xu, Baur:1989cm}. We can write
the amplitude as,

\begin{equation}
g(p_1^{A,\mu}) + g(p_2^{B,\nu}) \rightarrow g(-p_3^{C,\sigma}) + H(-p_5)
\end{equation}

If we define the partonic (with hats) kinematic variables in terms of
the Higgs momentum so that our differential cross-section can be written
in terms of the Higgs transverse momentum as $\hat{s}=(p_1+p_2)^2$,
$\hat{t}=(p_1+p_5)^2$, and $\hat{u}=(p_2+p_5)^2$. The matrix elements
take the symmetric form\cite{Dawson:1990zj, Djouadi:1991tk,
Kauffman:1991jt, Kauffman:cx, Kauffman:1993nv, Spira:1995rr,
Ravindran:2002dc, Field:2002pb}

\begin{align} \nonumber
|\mathcal{M}(gg\rightarrow gH)|^2 = &
 \frac{\as^3}{v^2} \frac{4N_c (N_c^2-1)}{9 \pi}
 \biggl( \frac{4\pi\mu_r^2}{m_\textrm{top}^2} \biggr)^{2\epsilon}
 \Gamma^2(1+\epsilon) \\
 \times & \biggl\{ \biggl[
 \frac{M_H^8+\hat{s}^4+\hat{t}^4+\hat{u}^4}{\hat{s}\hat{t}\hat{u}}
 \biggr] (1-2\epsilon)
 +\frac{\epsilon}{2} \biggl[
 \frac{(M_H^4+\hat{s}^2+\hat{t}^2+\hat{u}^2)^2}{\hat{s}\hat{t}\hat{u}}
 \biggr] \biggr\}.
\end{align}

Color and spin averaging gives an additional factor of
$1/256/(1-\epsilon)^2$. By pulling out the LO cross-section from the
expression, we can write the properly averaged matrix elements (where
the overbar corresponds to color and spin averaging)

\begin{align} \nonumber
|\overline{\mathcal{M}}(gg\rightarrow gH)|^2 =
 \hat{\sigma}_0^H \frac{\as}{\pi} \frac{N_c (N_c^2-1)}{1-\epsilon}
 \biggl\{ \biggl[
 \frac{M_H^8+\hat{s}^4+\hat{t}^4+\hat{u}^4}{\hat{s}\hat{t}\hat{u}}
 \biggr] (1-2\epsilon)
 +\frac{\epsilon}{2} \biggl[
 \frac{(M_H^4+\hat{s}^2+\hat{t}^2+\hat{u}^2)^2}{\hat{s}\hat{t}\hat{u}}
 \biggr] \biggr\}.
\end{align}

First, let us find the differential cross-section. The LO differential 
cross-section involves $2 \rightarrow 2$ kinematics and can be written 
with the $4-2\epsilon$ phase space dimensions as\cite{Kauffman:cx}

\begin{equation}
\frac{d\hat{\sigma}^H}{d\hat{t}} = \frac{1}{16\pi\hat{s}^2}
 \biggl( \frac{4 \pi \mu_r^2}{M_H^2} \biggr)^\epsilon
 \frac{1}{\Gamma(1-\epsilon)}
 \biggl( \frac{\hat{s} M_H^2}{\hat{u}\hat{t}} \biggr)^\epsilon
 |\overline{\mathcal{M}}|^2.
\end{equation}

As we will see in Sec.~\ref{resum}, it is the small $p_t$ behavior of
this expression that we will be interested in. If we insert the
expression $\hat{u}\hat{t} = \hat{s}p_t^2$ and drop the terms
proportional to $p_t$, we find an expression for the differential
cross-section in the small $p_t$ limit. Once we have changed variables,
we find the partonic differential cross-section in the small $p_t$
region (where we are suppressing the trivial rapidity dependence)

\begin{equation}
\label{diffDist}
\frac{d\hat{\sigma}^H}{dp_t^2} =
 \hat{\sigma}_0^H \frac{\as}{\pi}
 \biggl( \frac{4 \pi \mu_r^2}{M_H^2} \biggr)^\epsilon
 \frac{z}{(1-\epsilon)\Gamma(1-\epsilon)}
 \biggl( \frac{M_H^2}{p_t^2} \biggr)^\epsilon
 \biggl[ \ca \frac{1}{p_t^2} \ln \biggl( \frac{M_H^2}{p_t^2} \biggr) 
       - \beta_0 \frac{1}{p_t^2}
 \biggr].
\end{equation}

We will see that this representation will make it particularly simple to
extract the resummation coefficients $A^{(1)}_g$ and $B^{(1)}_g$ for the
differential distribution. These coefficients will be the same for the
total cross-section as well. Turning our attention back to the
expression we had for the matrix elements squared, we would like to
calculate the real corrections for the NLO total partonic cross-section.
The matrix elements for the real emission needs to be integrated and can 
be written as\cite{Dawson:1990zj}

\begin{equation}
\hat{\sigma} = \frac{1}{2 \hat{s}} \int \overline{|\mathcal{M}|}^2 
\, \textrm{dPS}_2 .
\end{equation}

For a $2 \rightarrow 2$ process, this can be done by introducing the
following parameterization for the angular integration. If we write the
scattering angle $\theta$ as $\cos\theta=2\omega -1$, this maps the 
$\theta$ integration to an $\omega$ integration between $0$ and $1$. We 
can express the kinematic invariants as follows
\begin{equation}
\hat{t} = -\hat{s}(1-z)(1-\omega), \quad \hat{u} = -\hat{s}(1-z)\omega,
\end{equation}
and do the phase space integrations using the following parameterization
\begin{equation}
\textrm{dPS}_2
     = \frac{1}{8\pi} \biggl( \frac{4\pi\mu_r^2}{\hat{s}} 
                      \biggr)^\epsilon
       \frac{(1-z)^{1-2\epsilon}}{\Gamma(1-\epsilon)}
       \int^1_0 \omega^{-\epsilon} (1-\omega)^{-\epsilon} d\omega
\end{equation}
which reduces the angular integration into repeated applications of
Euler's beta function integral, where additional integer powers of
$\omega$ and $(1-\omega)$ are introduced from the kinematic variables
$\hat{t}$ and $\hat{u}$,

\begin{equation}
\int^1_0 d\omega \, \omega^\alpha (1-\omega)^\beta =
\frac{\Gamma(1+\alpha)\Gamma(1+\beta)}{\Gamma(2+\alpha+\beta)}.
\end{equation}

Turning our attention back to the real emissions, we find the color and
spin averaged partonic total cross-section, after regulating the
singularity at $z=1$ with a plus distribution, can be written as

\begin{align} \nonumber
\hat{\sigma}_{\textrm{real}}^H = 
     \hat{\sigma}_0^H
     \biggl( \frac{\as}{\pi} \biggr)
  &  \biggl( \frac{4\pi\mu_r^2}{\hat{s}} \biggr)^{\epsilon}
     \Gamma(1+\epsilon) \\ \nonumber
     \times \ca \biggl\{ 
     \biggl[ & \frac{1}{\epsilon^2} + \frac{1}{\epsilon} + 1 - 2\z2 
     \biggr] \delta(1-z)
   - \frac{2z}{\epsilon} \biggl[ z \D0 + \frac{1-z}{z}
                                 + z(1-z)
     \biggr] - \frac{11}{6}(1-z)^3 \\ 
  &+ 2 \biggl[ 1+z^4+(1-z)^4 \biggr] \D1
   - 2 \biggl[ 1 - z + 2z^2 + z^4 \D0
     \biggr]
     \biggr\}
     + \mathcal{O}(\epsilon),
\end{align}
where the plus prescription is defined as usual
\begin{equation}
\int_0^1 dx \, f(x) [g(x)]_+ = 
\int_0^1 dx \, g(x) [f(x)-f(1)],
\end{equation}
and we have introduced the common abbreviation

\begin{equation}
\mathcal{D}_n(z) \equiv \biggl( \frac{\ln^n(1-z)}{1-z} \biggr)_\plus.
\end{equation}

\subsubsection{Virtual Corrections}
\label{scalarVirt}

Next, we must calculate the virtual contributions. There are two
diagrams with gluon loops (a gluon triangle and a four point incoming
state as can be seen in Ref.~\cite{Dawson:1990zj}), and can be
calculated directly. We can write the integrated virtual contribution
that contributes at $\as^3$ to the total partonic cross-section in the
same fashion as the real emissions\cite{Dawson:1990zj, Djouadi:1991tk,
Kauffman:1991jt, Kauffman:cx, Kauffman:1993nv, Spira:1995rr,
Ravindran:2002dc, Field:2002pb},

\begin{equation}
\hat{\sigma}_{\textrm{virt}} = \hat{\sigma}_0 
     \biggl( \frac{\as}{\pi} \biggr) 
     \biggl( \frac{4\pi\mu_r^2}{M_H^2} \biggr)^{\epsilon}
     \Gamma(1+\epsilon) \ca
     \biggl\{ \biggl[ - \frac{1}{\epsilon^2} - \frac{1}{\epsilon}
              + \frac{5}{6} + 4\z2 \biggr] \delta(1-z)
     \biggr\}.
\end{equation}

As expected, the $\epsilon^2$ singularities cancel between the real
emission and virtual graphs. It will turn out that the differences
between the fixed order results for the scalar and the pseudoscalar come
from different virtual corrections. These virtual contributions will be
needed to cancel some $\epsilon$ poles in the expression we will derive
to determine the process dependent resummation coefficients, in
particular $C^{(1)}_{gg}$ for the differential distribution.

\subsubsection{Total cross-section}
\label{scalarTot}

The remaining $1/\epsilon$ singularities must be removed to find the
partonic total cross-section. The poles are cancelled in the charge
(coupling) renormalization and the Altarelli-Parisi subtraction.
Understanding the charge renormalization tells us why is was so
important to have the $\mathcal{O}(\epsilon)$ terms of the lowest order
cross-section. We can see that the counter-term can be
written\cite{Dawson:1990zj, Djouadi:1991tk}

\begin{equation}
\hat{\sigma}_{\textrm{ren}} = (4Z_g)\hat{\sigma}_0, \quad
Z_g = - \frac{\as}{\epsilon}(4 \pi)^{\epsilon-1} \Gamma(1+\epsilon) 
\beta_0, \quad
\beta_0 = \frac{11}{6}\Blue{\ca} - \frac{2}{3}\Blue{\nf\tr},
\end{equation}
where $\nf=5$ since the top quark has been integrated out. These 
equations hold with the $\overline{\textrm{MS}}$ renormalization 
conditions. 

The Altarelli-Parisi subtraction factors out the soft and collinear
singularities into the PDFs much like the factorization process
separates the short and long distance physics in hadron-hadron
scattering. This cancels the rest of the $1/\epsilon$ poles and gives us
the final expression for the total cross-section

\begin{align} \nonumber
\label{Hnlo}
\hat{\sigma}_{\textrm{NLO}}(gg \rightarrow H + X) = 
 \hat{\sigma}_0^H 
 \biggl\{ \delta(1-z) + \biggl( \frac{\as}{\pi} \biggr) &\biggl[ 
  \biggl( \frac{11}{6}\ca + 2\ca\z2 \biggr) \delta(1-z) 
        - \frac{11}{6} \ca (1-z)^3 \\
        + 2\ca \biggl[ 1 + z^4 + (1-z)^4 \biggr]\D1
& + 2\ca \biggr( z^2\D0 + (1-z) + z^2(1-z) \biggr) 
  \ln\frac{M^2_H}{z\mu^2} \biggr] \biggr\} + \mathcal{O}(\epsilon).
\end{align}

Our expressions for the resummation coefficients show the their full
color dependence. It is sufficient to notice at this stage that the term
proportional to the $\delta(1-z)$ in the correction can be evaluated as
$11/2 + \pi^2$ when we use

\begin{equation}
\Blue{\ca} = \nc, \quad \Blue{\cf} = \frac{\n2c-1}{2\nc}, \quad 
\Blue{\tr} = \frac{1}{2}.
\end{equation}

\subsection{Pseudoscalar Higgs Matrix Elements}
\label{pseudoME}

There are many reasons for our primary interest to be the pseudoscalar
Higgs boson. In the MSSM, the exact roles the top and bottom quarks play
in the differential cross-section is complicated\cite{Field:2003yy}. 
However, in much of the parameter space in the MSSM, the cross-section
for the pseudoscalar is larger than the lightest scalar Higgs boson in
the MSSM. If supersymmetry does exist in nature, the pseudoscalar Higgs
may be the first Higgs boson discovered due to its larger cross-section. 
If supersymmetry becomes important only at very high scales, then seeing
a pseudoscalar Higgs would be the first evidence of supersymmetry in
nature. This leads us to investigate in detail pseudoscalar resummation.

We should also mention that because of the importance of the bottom
quark in calculations involving the MSSM pseudoscalar (and the lightest
scalar Higgs in the MSSM as well) when the parameter $\tan\beta$ is
large is systematically ignored in these calculations. To remedy this
situation, one would have to calculate the resummation coefficients in
the full theory. In principle, these coefficients can be extracted from
Ref.~\cite{Spira:1995rr}. However, these results have not been
published.

The difference in the lowest order (LO) partonic cross-section of the 
pseudoscalar Higgs boson and the scalar Higgs boson can be traced to the 
difference in the effective couplings $g_H$ and $g_{A^0}$ and a factor 
of $9/4$. This can be written as\cite{Kauffman:1993nv}

\begin{align}
\hat{\sigma}_0^{A^0} (gg\rightarrow A^0) 
            & = \frac{9}{4} \hat{\sigma}_0^H
                 (gg\rightarrow H) \\ \nonumber
            & = \biggl( \frac{\as}{\pi} \biggr)^2
                \frac{\pi}{256 v^2}
                \biggl( \frac{4\pi\mu_r^2}{m_\textrm{top}^2} 
                \biggr)^{2\epsilon}
                \frac{\Gamma^2(1+\epsilon)}
                     {1-\epsilon} \, \delta(1-z).
\end{align}

Upon the expansion of the $\mathcal{O}(\epsilon)$ terms, we see they do 
not effect the final answer as expected at the lowest order. 

\subsubsection{Radiative Corrections}
\label{pseudoRad}

The matrix elements for the production of the pseudoscalar in the HQET
are slightly more complicated in $d=4-2\epsilon$ dimensions due to the
presence of the intrinsically $4$-dimensional Levi-Civita tensor in the
Feynman rules\cite{Kauffman:1993nv} coming from the $\tilde{G}^{a,\mu
\nu}$ term in the effective Lagrangian. There are several conventions
for handling this problem\cite{'tHooft:fi, Akyeampong:xi,
Akyeampong:1973vk, Akyeampong:1973vj}. We have chosen the scheme defined
in Ref.~\cite{Akyeampong:xi, Akyeampong:1973vk, Akyeampong:1973vj}.

For the radiative corrections, we can separate all the vectors into $4$-
and $(d-4)$-dimensional components. We label the $(d-4)$-dimensional
components of the vectors with a twiddle. We can take the incoming
momentum to be $4$-dimensional as a convenient choice of frame, which
simplifies the results considerably\cite{Kauffman:1993nv}.

The (un-averaged) matrix elements can be written,

\begin{align} \nonumber
|\mathcal{M}(gg\rightarrow gA^0)|^2 = &
 \frac{\as^3}{v^2} \frac{N_c (N_c^2-1)}{\pi}
 \biggl( \frac{4\pi\mu_r^2}{m_\textrm{top}^2} \biggr)^{2\epsilon}
 \Gamma^2(1+\epsilon) \\
 \times & \biggl\{ \biggl[
 \frac{M_{A^0}^8+\hat{s}^4+\hat{t}^4+\hat{u}^4}{\hat{s}\hat{t}\hat{u}}
 \biggr] +
 \frac{ 2\hat{s}(\hat{t}^2+\hat{u}^2)
                (\tilde{p}_3 \! \cdot \! \tilde{p}_3
                 \hat{s}-\hat{t}\hat{u}\epsilon)
      }
      {\hat{t}^2\hat{u}^2}
 \biggr\}.
\end{align}

We can see that the $\epsilon \rightarrow 0$ corrections are identical
in the scalar and pseudoscalar case for the real emissions. We can also
see that the residual difference is not only proportional to $\epsilon$,
but rather proportional to $\epsilon$ and the $(d-4)$-dimensional
component of the $p_3$ vector, which vanishes in the $4$-dimensional
limit.

From this analysis, we can see that the real part of the differential
cross-section for the pseudoscalar Higgs boson will be identical to the
scalar case in Equation~(\ref{diffDist}) with a $9/4$ difference in
normalization.

\subsubsection{Virtual Corrections}
\label{pseudoVirt}

The virtual corrections to the pseudoscalar have the same diagrams as
the scalar at this order, but there is a slight difference in the
result. This difference is due to the fact that the diagram with the
four point gluon vertex vanishes due to the antisymmetry of the
$ggA^{0}$ vertex in the effective theory. The integrated result is

\begin{equation}
\hat{\sigma}_{\textrm{virt}}^{A^0} = \hat{\sigma}_0^{A^0}
     \biggl( \frac{\as}{\pi} \biggr) 
     \biggl( \frac{4\pi\mu_r^2}{M_H^2} \biggr)^{\epsilon}
     \Gamma(1+\epsilon) \ca
     \biggl\{ \biggl[ - \frac{1}{\epsilon^2} - \frac{1}{\epsilon}
              + 2 + 4\z2 \biggr] \delta(1-z)
     \biggr\}.
\end{equation}

We can see that although the pole terms are the same, the finite terms 
have changed slightly because of the ``missing'' diagram.

\subsubsection{Total cross-section}
\label{pseudoTot}

When we combine our pseudoscalar results, we find that the total
partonic cross-section that is identical to the scalar case with the
exception of the small numeric difference in the $\delta(1-z)$ term. The
expression changes from $11/6 \ca + 2 \ca \z2 \rightarrow 2 \ca + 2 \ca
\z2$ in the pseudoscalar case. Here we see the factor $11/2 \rightarrow
6$, which would seem to be a small difference, but it is mostly a
coincidence of the $\textrm{SU}_3$ Casimir invariants. 

The partonic cross-section can be written as

\begin{align} \nonumber
\label{Anlo}
\hat{\sigma}^{A^0}_{\textrm{NLO}}(gg \rightarrow A^0 + X) = 
 \hat{\sigma}_0^{A^0} 
 \biggl\{ \delta(1-z) + \biggl( \frac{\as}{\pi} & \biggr)
 \biggl[ \biggl( 2\ca + 2\ca\z2 \biggr) \delta(1-z) 
       - \frac{11}{6} \ca (1-z)^3 \\
  + 2 \ca \biggl[ 1 + z^4 + (1-z)^4 \biggr] \mathcal{D}_1(z)
& + 2 \ca \biggr( z^2\D0 + (1-z) + z^2(1-z) \biggr) 
  \ln\frac{M^2_H}{z\mu^2} \biggr] \biggr\} + \mathcal{O}(\epsilon).
\end{align}

Now that we have expressions for the total partonic cross-sections for
both the scalar and the pseudoscalar Higgs boson we see that the only
difference between the two lies in the correction proportional to
$\delta(1-z)$ at NLO and in the normalization. This difference in the
$\delta(1-z)$ factors is numerically small and is $\as$ suppressed,
leaving us to believe that the primary difference is going to be factor
of $9/4$ in the LO partonic cross-sections.

\section{Resummation}
\label{resum}

To introduce the machinery behind resummation\cite{Collins:1981uk,
Collins:va, Collins:1984kg}, we need to define the hadronic
cross-section. This is the convolution of parton distributions functions
(PDFs) with the partonic cross-section

\begin{equation}
\sigma(S,M_\Phi^2) = \sum_{a,b} 
\int_{x_{1,\textrm{min}}}^1 \! \! dx_1
\int_{x_{2,\textrm{min}}}^1 \! \! dx_2 \,
f_{a/h_{1}}(x_1,\mu_f) \, f_{b/h_{2}}(x_2,\mu_f) \,
\int_0^1 dz \,
z \, \hat{\sigma}_0^\Phi \Delta_{ab \rightarrow \Phi}(z,\mu_r,\mu_f)
\end{equation}
where $\mu_r$ and $\mu_f$ are the renormalization and factorization
scales respectively, and $f_{a/h_{1}}\!$ is the parton distribution
function for finding a parton $a$ in hadron $h_1$. We must also remember
that there is a $\delta(1-z)$ in the definition of the LO cross-section
$\hat{\sigma}_0^\Phi$. The minimum partonic energy fraction
$x_{(1,2),\textrm{min}}$ is defined so that there is enough center of
momentum energy to create the desired final state particles. A similar
equation can be written for the differential distribution.

Implicitly, the partonic cross-section contains logarithmic
corrections that are formally singular at threshold ($z \rightarrow 1$).
The differential cross-section contains corrections that are singular as
the transverse momentum of the Higgs particle vanishes. They can be
written in the form

\begin{equation}
\textrm{threshold} \sim \as^n\frac{\ln^{2n-1} (1-z)}{(1-z)},
\quad
\textrm{recoil}    \sim \frac{\as^n}{p_t^2} 
                        \ln^{2n-1}\frac{M_\Phi^2}{p_t^2},
\end{equation}
and various powers of these combinations. It can be seen then that the
normal fixed order cross-section calculation diverges (in one direction
or the other) at small $p_t$ due to large logarithms, and therefore it
is not reliable in this region. The systematic way of handling these
formally divergent terms at small $p_t$ is known as resummation. Because
of this divergent behavior, one is not usually able to integrate the
differential cross-section all the way down to $p_t=0$ or to reliably
understand the differential cross-section in the experimentally
interesting small $p_t$ region. Resummation coefficients can be
determined for both differential distributions and total-cross sections
to address this problem.

\subsection{Formalism}
\label{formalism}

The resummation formalism allows the small $p_t$ cross-section to be
written as a power series in both universal and process dependent
coefficients. We write the resummed differential cross-section for a
$c\bar{c} \rightarrow \Phi$ process (where $c$ in this case represents a
gluon or a quark) 

\begin{equation}
\label{resumEqn}
\frac{d\sigma^{\textrm{resum}}}{dp_t^2 \, dy \, d\phi} 
= \sum_{a,b} \int_{x_{1,\textrm{min}}}^1 \!\!\! dx_1
\int_{x_{2,\textrm{min}}}^1 \!\!\! dx_2 
\int_0^\infty db \, \frac{b}{2} J_0(bp_t) \,
f_{a/h_1}\!(x_1,b_0/b) \, f_{b/h_2}\!(x_2,b_0/b) \,
\frac{S}{Q^2} W_{ab}(x_1 x_2 S;Q,b,\phi),
\end{equation}

\begin{equation}
W_{ab}(s;Q,b,\phi) = \sum_c \int_0^1 dz_1 \int_0^1 dz_2 \,
\bar{C}_{ca}(\as(b_0/b),z_1) \, \bar{C}_{\bar{c}b}(\as(b_0/b),z_2) \,
\delta(Q^2-z_1 z_2 s) \, \frac{d\sigma_{\bar{c}c}^{LO}}{d\phi}
\, S_c(Q,b),
\end{equation}
where the Higgs mass $M_\Phi^2 = Q^2$, $d\phi$ is the phase space of the
system under consideration, and $\hat{\sigma}^{(LO)}_{c\bar{c}}$ is the
lowest order cross-section with a $c\bar{c}$ initial state which is
therefore defined at $p_t=0$. The constant $b_0$ is written in terms of
the Euler-Mascheroni constant $\gamma_E=0.57721\ldots$ as
$b_0=2e^{-\gamma_E}$. In the resummation formalism only the $gg$ and
$q\bar{q}$ initial states are needed to determine the hadronic
differential distribution at small $p_t$. The $qg$ initial states are
accounted for in the cross terms in the convolution.  The coefficients
$C_{ab}$ are process dependent and can be written as power series to
be described below. $J_0(bp_t)$ is the first order Bessel function. The
Sudakov form factor $S_c$, which makes the integration over the Bessel
function convergent, can be written as

\begin{equation}
S_c(Q,b) = \exp \biggl\{
- \int^{Q^2}_{b_0^2/b^2} \frac{dq^2}{q^2} \biggl[
A_c(\as(q)) \ln\frac{Q^2}{q^2} + B_c(\as(q))
\biggr] \biggr\}.
\end{equation}
The coefficient functions $A_c$, $B_c$, and $C_{ab}$ can be written as
power series in $\as$ as
\begin{equation}
A_c(\as) = \sum_{n=1}^{\infty} 
                \biggl( \frac{\as}{\pi} \biggr)^n A_c^{(n)}, \quad
B_c(\as) = \sum_{n=1}^{\infty} 
                \biggl( \frac{\as}{\pi} \biggr)^n B_c^{(n)},
\end{equation}
\begin{equation}
\bar{C}_{ab}(\as,z) = \delta_{ab}\delta(1-z) +
                      \sum_{n=1}^{\infty} \biggl( \frac{\as}{\pi} 
                      \biggr)^n \bar{C}_{ab}^{(n)}(z).
\end{equation}

The $A^{(1)}_c$, $A^{(2)}_c$, and $B^{(1)}_c$ coefficients have been
shown to be universal. There are several conventions in the literature
as to whether to expand in terms of $\as/\pi$ or $\as/2\pi$ (or even
$\as/4\pi$ in Ref.~\cite{Ravindran:2003um}). We have chosen to expand in
$\as/\pi$. It would seem that several typos exist in the literature due
to this numeric expansion factor. We have derived the previously unknown
coefficients $B^{(2)}_g$, $C^{(1)}_{gg}$ and $C^{(2)}_{gg}$ for
pseudoscalar Higgs production for the total cross-section and the
$\bar{C}^{(1)}_{gg}$ for the differential cross-section resummation
given below. Here we must stop to address a question of notation. It is
unfortunate that we have the same notation for the resummation
coefficients for both the $p_t$ resummation and the total cross-section
resummation. It would be convenient to use a calligraphic font for the
$p_t$ coefficients, but several authors have used this font in other
contexts dealing with resummation. Therefore, we will put bars over the
resummation coefficients for differential cross-sections even if they
are identical to the coefficients for the total cross-section
resummation.

To determine the $\bar{C}^{(n)}$ coefficients for the differential
cross-section, one must understand the meaning of the resummation
formula. We can expand Equation~(\ref{resumEqn}) order by order in $\as$
and compare to the perturbative calculation to read off the
coefficients\cite{deFlorian:2001zd}. To extract the coefficients from
the perturbative results, we need to integrate the differential
cross-section around $p_t=0$ paying careful attention to the use of the
Altarelli-Parisi splitting functions near $p_t = 0$ as follows

\begin{equation}
\Delta\hat{\sigma} = \int_0^{q_t^2}
 dp_t^2 \frac{d\hat{\sigma}}{dp_t^2}.
\end{equation}

This expression will contain $\epsilon$ poles and virtual corrections at
the next order will be needed to be added to find a finite expression. 
This is demonstrated later in this paper. One should also be careful to
use this formula with other quantities that show the same rapidity
dependence. When we expand to $\mathcal{O}(\as)$, we can see that the
NLL coefficients emerge as follows for a $c\bar{c}$ initial state

\begin{equation}
\Delta\hat{\sigma}_{c\bar{c}} = 1 +
\frac{\as}{\pi} \biggl[ 
 - \frac{\bar{A}^{(1)}_c}{2} \ln^2 \biggl( \frac{M_\Phi^2}{q_t^2} 
                                   \biggr)
 - \bar{B}^{(1)}_c \ln \biggl( \frac{M_\Phi^2}{q_t^2} \biggr)
 + 2 \bar{C}^{(1)}_{c\bar{c}}
                \biggr].
\end{equation}

In principle, it is possible to continue this process to higher orders
to obtain the needed coefficients for the differential cross-section. 
The NLO corrections to the differential cross-section are
known\cite{Glosser:2002gm, Ravindran:2002dc}, however the NNLO
differential cross-section for Higgs production is currently unknown.
However, the total cross-section is known to NNLO, so the resummation
coefficients for the total cross-section can be determined to NNLL. The
NLO differential cross-section in Ref.~\cite{Glosser:2002gm} has been
written in terms of the $p_t$ of the Higgs boson for the scalar case,
and could in principle be used in part to extract the NNLO process
dependent $\bar{C}^{(2)}_{gg}$ coefficient for the differential
cross-section for the scalar Higgs boson, advancing the resummed
expressions ahead of the fixed order calculation. This work has not yet
been completed.

\subsection{Matching}
\label{matching}

The resummation formalism is valid in the small $p_t$ region. Fixed
order perturbation theory is valid at moderate $p_t$ where there are no
large logarithms. The process of matching allows for a smooth
transition between the two regions. The procedure is described in great
detail and clarity in Ref.~\cite{Arnold:1990yk}.

One can write the differential cross-section as the sum of three terms
\begin{equation}
\frac{d\sigma}{dp_t^2 \, dy} =
\frac{d\sigma^{\textrm{resum}}}{dp_t^2 \, dy} +
\frac{d\sigma^{\textrm{pert}}}{dp_t^2 \, dy} -
\frac{d\sigma^{\textrm{asym}}}{dp_t^2 \, dy}
\end{equation}

This equation is easy to understand. At low $p_t$, we have the resummed
contribution since the latter contributions cancel. At high $p_t$ we
have the perturbative contribution when the resummed and asymptotic
cancel. At small $p_t$ we remove the terms from the perturbative
expansion that are asymptotically divergent like $1/p_t^2$. This allows
for a smooth transition between the two regions at all values of $p_t$.
However, extracting the divergent pieces can be quite difficult
analytically as one must express the differential cross-section in terms
of $p_t$ order by order. For a $2 \rightarrow 1$ process, the first
order corrections have $2 \rightarrow 2$ kinematics and this is
relatively simple, but becomes more intractable for the higher order
corrections.

In this paper, we are interested in the new coefficient functions, and
determining where the distributions peak for the different colliders, so
this treatment will be ignored. However, we will display the
perturbative differential cross-section to guide the eye on what the
transition must look like.

\subsection{Higgs Resummation}
\label{higgsResum}

One of the interesting facets for Higgs production is that there is only
a $gg$ initial state for this process in the HQET at order $\as^3$, so
the other terms (a $q\bar{q}$ initial state) are zero explicitly.
Without getting too far ahead of our discussion, we can see that the
Mellin moments of the $q\bar{q}$ corrections at order $\as^4$ strictly
vanish on threshold due to the fact that there is no $q\bar{q}$ initial
state at lowest order. This makes it possible to work in $z$-space with
little additional effort due to the presence of the $\delta(1-z)$ terms
in the $C^{(n)}_{gg}$ coefficients, which makes the convolution with the
PDFs trivial.

An additional complication arises from evaluating the parton
distribution functions at very low scales during the convolution. This
is solved by what is known as the $b_\star$
prescription\cite{Arnold:1990yk, Davies:1984hs}. Here the $b$ parameter
is replaced by $b_\star$ that has an infrared cut-off $b_{\text{max}}$
so that as $b$ becomes large, $b_\star \rightarrow b_{\text{max}}$, and
the fraction $b_0/b$ in Equation~(\ref{resumEqn}) never leaves the
perturbative regime of the parton distribution function. Over the rest of
the range $b_\star \approx b$. This can be achieved by in the following
parametrization

\begin{equation}
b_\star = \frac{b}{\sqrt{1+b^2/b^2_{\textrm{max}}}}.
\end{equation}

This construction may seem a little artificial, but it allows for the 
numeric integration of our differential cross-section and allows us to 
use what is known to make reasonable calculations.

In these calculations, we have set $b_{\text{max}} = (2$~GeV$)^{-1}$. 
This is mostly determined by the limits of applicability for the PDFs
implemented to obtain the hadronic cross-section. There are other
non-perturbative correction factors that are
employed\cite{Berger:2002ut, Berger:2003pd, Arnold:1990yk}, but as no
data is available yet we have not included these factors in our
analysis. 

It is thus possible to determine the unknown coefficients to a given
order in the resummation and preform the resummed calculation. The known
coefficients for scalar Higgs production will be given later. To
leading-log (LL) accuracy, only the $A^{(1)}_c$ term is needed. At
next-to-leading log (NLL) accuracy one needs the $A^{(2)}_c$,
$B^{(1)}_c$, and $C^{(1)}_{ab}$ coefficients. The state of the art
currently is NNLL where the $A^{(3)}_c$, $B^{(2)}_c$, and $C^{(2)}_{ab}$
coefficients are needed\cite{deFlorian:2001zd}. For a scalar Higgs,
these terms have been recently calculated but are missing for a
pseudoscalar Higgs.  The $A^{(3)}_g$ term can now be determined thanks
to the excellent recent work on the three-loop splitting
functions\cite{Vogt:2004mw, Moch:2004pa}. Previously, only a numeric
estimate was available\cite{Vogt:2000ci}.

Let us begin by extracting the process dependent $C_{ab}$ coefficients
for the total cross-section. To extract the formally divergent pieces of
the cross-section, consider the Mellin transform of the hadronic
cross-section, $\sigma_N(M_\Phi^2)$. The $N-$moments in Mellin space are
defined as

\begin{equation}
\sigma_N(M_\Phi^2) \equiv \int_0^1 dz \, z^{N-1} \, \sigma(z,M_\Phi^2)
\end{equation}
The advantage of transforming to Mellin space is that the limit $z
\rightarrow 1$ corresponds to the limit of $N \rightarrow \infty$. This 
allows for a systematic way of extracting the divergent terms, which 
diverge as $\ln(N)$ in Mellin space.

Before continuing, we should comment on which initial state channels
contribute. In evaluating the the $C_{ab}$ coefficients in the $N
\rightarrow \infty$ limit we see that only the $gg$ channel has finite
contributions, all the other channels have Mellin moments are strictly
zero on threshold. We could also see that in the HQET there are no
$q\bar{q}$ or $qg$ initial state that contribute at this order to the
cross-section at $p_t=0$. Although we have set up our formalism for the
sum of several channels, we will now consider only the $gg$ initial
state channel.

The Mellin moments of the fixed order corrections allow us to determine
the process dependent total cross-section $C^{(n)}_{gg}$ coefficients in
a simple way. We find the Mellin moments of the fixed order corrections,
$\Delta^{(n)}_{ab\rightarrow \Phi}$, with the package \textsc{harmpol}
in \textsc{form}\cite{Vermaseren:2000nd}. Some diverge as $\ln(N)$, most
tend to zero as $N \rightarrow \infty$, and some finite pieces are left
over. In this way, we can separate the formally divergent pieces from
the finite contributions on threshold. With this we can identify the
$p_t$ divergent terms with the $\ln(N)$ divergent terms in the Mellin
moment.  We choose to absorb the extra powers of $\gamma_E$ into our
definition of $\tilde{N} = N e^{\gamma_E}$ so that we do not have
spurious factors of $\gamma_E$ in our expressions. This seems to be
appropriate as in the $\overline{\textrm{MS}}$ scheme the factors of
$\gamma_E$ are also absorbed.  This being noted, we will continue to
write our terms as $\ln(N)$ with no factors of $\gamma_E$.

\subsubsection{Next-to-leading-log Differential 
Cross-section Coefficients}
\label{nllDiff}

We are interested in determining the $C^{(1)}_{gg}$ coefficient for the 
scalar and pseudoscalar Higgs boson. First, let us integrate the 
differential cross-section for the scalar around $p_t=0$. We will label 
this contribution `real' to note that this integral is similar to the 
real emission corrections to the total cross-section. We find

\begin{equation}
\Delta\hat{\sigma}^{\textrm{real}} = \hat{\sigma}_0^H
 \biggl( \frac{4\pi\mu_r^2}{M_H^2} \biggr)^\epsilon
 z \Gamma(1+\epsilon)(1+\epsilon) \frac{\as}{\pi}
 \biggl[ \frac{\ca}{\epsilon^2} + \frac{\beta_0}{\epsilon}
        -\frac{\ca}{2} \ln^2 \biggl( \frac{M_H^2}{q_t^2} \biggr)
        +\beta_0 \ln \biggl( \frac{M_H^2}{q_t^2} \biggr)
        + \ca - \ca\z2
 \biggr].
\end{equation}

We have to add the total partonic cross-section virtual correction to
this expression to cancel the $\epsilon$ poles. The pole proportional to
the $\beta_0$ gets renormalized into the coupling like in the total
cross-section calculation. Once these two expression are added together
we find

\begin{equation}
\Delta\hat{\sigma} = \hat{\sigma}_0^H z \biggl[
1 + \frac{\as}{\pi} \biggl(
-\frac{\ca}{2} \ln^2 \biggl( \frac{M_H^2}{q_t^2} \biggr)
+\beta_0 \ln \biggl( \frac{M_H^2}{q_t^2} \biggr)
+ \ca + \frac{5}{6}\ca + 3\ca\z2
\biggl) \biggr].
\end{equation}

The coefficients can now be read off and agree with the 
literature\cite{Kauffman:cx}

\begin{equation}
\bar{A}^{(1),H}_g    = \Blue{\ca}, \quad
\bar{B}^{(1),H}_g    = - \beta_0 = 
                       - \biggl( \frac{11}{6}\Blue{\ca} - 
                         \frac{2}{3}\Blue{\nf\tr} \biggr), \quad
\bar{C}^{(1),H}_{gg} = \frac{11}{12}\Blue{\ca} + 
                       \frac{3}{2}\Blue{\ca}\z2.
\end{equation}

As noted earlier, the pseudoscalar Higgs has different virtual
corrections from the scalar Higgs boson. This changes the
$\bar{C}^{(1)}_{gg}$ coefficient for the pseudoscalar to

\begin{equation}
\bar{C}^{(1),A^0}_{gg} = \Blue{\ca} + \frac{3}{2}\Blue{\ca}\z2.
\end{equation}

The pseudoscalar $\bar{C}^{(1)}_{gg}$ coefficient is larger that the
scalar coefficient by a factor of $1/12\ca$. This is a small numeric
difference, and the NNLL coefficient have not been extracted for the
differential distribution, although as we will see in the next section
we might expect a larger difference to appear in the differential
$\bar{C}^{(2)}_{gg}$ coefficient based on the differences in the
$C^{(2)}_{gg}$ coefficients for the total cross-section.

\subsubsection{Next-to-leading-log Total Cross-section Coefficients}
\label{nllTotal}

Exact expressions for the fixed order NLO corrections to scalar and
pseudoscalar Higgs production have been in the literature for some time. 
Leaving aside the Sudakov terms ($A^{(n)}$ and $B^{(n)}$), let us
examine our expressions for the fixed order corrections to the partonic
cross-section. We see that the NLO corrections have organized themselves
in terms of constant pieces proportional to $\delta(1-z)$ from the soft
and virtual corrections and additional logarithmic corrections. The
Mellin moment of the $\delta(1-z)$ is simply

\begin{equation}
\int_0^1 dz \, z^{N-1} \, \delta(1-z) = 1.
\end{equation}

So it is easy to see that all the constant terms proportional to
$\delta(1-z)$ contribute to the $C_{gg}^{(1)}$ term. Beyond these terms,
the Mellin moments of the logarithmic corrections in the limit $N
\rightarrow \infty$ can have finite pieces that also contribute. Once 
the expression for the correction term has been transformed into Mellin 
space, there are no terms proportional to $\delta(1-z)$, but are only 
constant terms.

In presenting the expressions for the $C_{gg}^{(n)}$ terms, we mix the
notation somewhat to allow the reader to see all the different
contributions. We keep the $\ln(N)$ pieces that are formally divergent,
we separate out the terms that were initially proportional to the
$\delta(1-z)$ for convenience, and we include the terms proportional to
$\ln(M^2_\Phi/\mu^2)$ for completeness. We have set $\mu = \mu_r =
\mu_f$ for simplicity. The finite pieces compose the $C_{gg}^{(n)}$ 
coefficients.

In the case of the scalar and pseudoscalar
\begin{equation}
\Delta^{(1),H}_{N,gg} = 
\lim_{N \rightarrow \infty} \int_0^1 dz \, z^{N-1} 
                            \Delta^{(1)}_{gg \rightarrow H} = 
                            \Blue{\ca} \ln^2(N)
                            - 2 \Blue{\ca} 
                              \ln\frac{M^2_H}{\mu^2} \ln(N) +
                            \biggl[ \frac{11}{6} \Blue{\ca} +
                                               2 \Blue{\ca} \z2
                            \biggr]
                            + 2 \Blue{\ca} \z2,
\end{equation}
\begin{equation}
\Delta^{(1),A^{0}}_{N,gg} = 
\lim_{N \rightarrow \infty} \int_0^1 dz \, z^{N-1} 
                            \Delta^{(1)}_{gg \rightarrow A^{0}} = 
                            \Blue{\ca} \ln^2(N)
                            - 2 \Blue{\ca} 
                              \ln\frac{M^2_{A^0}}{\mu^2} \ln(N) +
                            \biggl[ 2 \Blue{\ca} + 
                                    2 \Blue{\ca} \z2 \biggr]
                            + 2 \Blue{\ca} \z2 ,
\end{equation}
where the terms in the square brackets are the terms that were
proportional to the delta function in the expression for the NLO
correction in Equations~(\ref{Hnlo}) and (\ref{Anlo}). We have used the
convention of absorbing the extra factors of $\gamma_E$ that appear in
other expressions for the $C_{gg}^{(n)}$ coefficients.

\subsubsection{Next-to-next-to-leading-log Total 
Cross-section Coefficients}
\label{nnllTotal}

Exact expressions for the fixed order NNLO corrections to scalar and
pseudoscalar Higgs production are known. The NNLO corrections to
inclusive Higgs production have been explicitly calculated and presented
for the scalar\cite{Harlander:2001is, Ravindran:2003um} and the
pseudoscalar\cite{Harlander:2002vv, Ravindran:2003um}. Although the the
color factors have been evaluated in \cite{Harlander:2001is,
Harlander:2002vv}, they were found to agree perfectly with
\cite{Ravindran:2003um} once the color factors were evaluated.

This allowed for the determination of the $C^{(2)}_{gg}$ coefficient for
both the scalar and pseudoscalar. The scalar result was compared with
the literature value\cite{Catani:2003zt} and was found to be in prefect
agreement once the factorization and renormalization scales were set
equal to one another and the spurious factors of $\gamma_E$ were
absorbed. For completeness, we present the full expression, leaving the
color dependence intact and showing the $\ln(N)$ contributions. 

\begin{align} \nonumber
\Delta_{N,gg}^{(2),H}(z) = &
   \biggl[ 2 \Blue{\caca} \biggl] \Red{\ln^4(N)}
+  \biggl[ \Blue{\caca} \biggl( \frac{11}{9} - 4 
\ln\frac{M^2_H}{\mu^2} \biggr)
- \frac{4}{9} \Blue{\nf \ca \tr} \biggr] \Red{\ln^3(N)} \\ \nonumber
+& \biggl[ \Blue{\caca}
\biggl( \frac{157}{18} + 7\z2 - \frac{11}{6} \ln\frac{M^2_H}{\mu^2} 
+ 2 \ln^2\frac{M^2_H}{\mu^2} \biggr) - 3 \Blue{\ca \cf}
- \Blue{\nf \ca \tr} \biggl( \frac{10}{9} - \frac{2}{3} 
\ln\frac{M^2_H}{\mu^2} \biggr)
\biggr] \Red{\ln^2(N)} \\ \nonumber
+& \biggl[
\Blue{\caca} \biggl( \frac{101}{27} - \frac{7}{2}\z3 
                   - \biggl( 7\z2 + \frac{157}{18} \biggr) 
                     \ln\frac{M^2_H}{\mu^2} 
                   + \frac{11}{12}\ln^2\frac{M^2_H}{\mu^2}
              \biggr)
+ 3 \Blue{\ca \cf} \ln\frac{M^2_H}{\mu^2} \\ \nonumber
&- \Blue{\nf \ca \tr}
                \biggl( \frac{28}{27}-\frac{10}{9}\ln\frac{M^2_H}{\mu^2}
                      + \frac{1}{3}\ln^2\frac{M^2_H}{\mu^2}
              \biggr)
\biggr] \Red{\ln(N)} \\ \nonumber
+& \biggl\{
\Blue{\caca}
      \biggl( \frac{3187}{288} + \frac{157}{18}\z2 - 
              \frac{1}{20}\zz2 - \frac{55}{12}\z3 +
              \frac{7}{8}\ln\frac{\mu^2}{m_{\textrm{top}}^2}
            - \biggl[
              \frac{3}{2} + \frac{11}{6}\z2 - \frac{19}{2}\z3
              \biggr] \ln\frac{M^2_H}{\mu^2}
            - 2\z2 \ln^2\frac{M^2_H}{\mu^2}
      \biggr) \\ \nonumber
&+ \frac{9}{4} \Blue{\cfcf} - \frac{1}{6} \Blue{\cf \tr} - 
                \frac{5}{48}\Blue{\ca \tr}
 - \Blue{\ca \cf}
           \biggl( \frac{145}{24} + 3\z2 + 
                   \frac{11}{8}\ln\frac{\mu^2}{m_{\textrm{top}}^2}
           \biggr) \\ \nonumber
&- \Blue{\nf \tr}
           \biggl( \Blue{\ca}
                       \biggl( \frac{1153}{216} +
                               \frac{10}{9}\z2 -
                               \frac{5}{9}\z3
                       \biggr)
                 + \Blue{\cf}
                       \biggl( \frac{3}{8} -
                               \ln\frac{\mu^2}{m_{\textrm{top}}^2}
                       \biggr)
 - \biggl[ \Blue{\ca} - \frac{2}{3} \z2 \Blue{\ca} - 
                        \frac{1}{2} \Blue{\cf}
   \biggr] \ln\frac{M^2_H}{\mu^2}
           \biggr)
\biggr\} \\ \nonumber
+& \Blue{\caca} \biggl[ \frac{157}{18} \z2 + \frac{29}{5} \zz2 +
           \frac{22}{9} \z3 -
           \biggl( \frac{11}{6} \z2 + 8 \z3
           \biggr)\ln\frac{M^2_H}{\mu^2} + 2\z2 
           \ln^2\frac{M^2_H}{\mu^2}
                \biggr] \\
-& \Blue{\ca \cf} \biggl[ 3 \z2
                  \biggr]
 - \Blue{\nf \tr \ca} \biggl[ \frac{10}{9} \z2 +
                              \frac{8}{9} \z3 -
                              \frac{2}{3} \z2
                              \ln\frac{M^2_H}{\mu^2}
                      \biggr]
\end{align}

The term in curly brackets all by itself was the piece proportional to
$\delta(1-z)$ in the original NNLO correction. The NNLO correction used
as input for the Mellin moment can be found in
Ref.~\cite{Ravindran:2003um}. One should carefully note that there are
terms in the coefficient that are proportional to $\tr$, but not
$\nf\tr$. These terms come from the higher order corrections to the
$g_H$ effective coupling where only the top quark is included in the
derivation of the corrections\cite{Chetyrkin:1997sg, Kramer:1996iq}. One
of our novel results is the $C^{(2)}_{gg}$ factor of the pseudoscalar,
which is very similar to the scalar case. The difference between the two
coefficients can be written as

\begin{align} \nonumber
\Delta_{N,gg}^{(2),A^0}-\Delta_{N,gg}^{(2),H} & =
         \biggl[
                 \frac{\Blue{\ca}}{4} 
                 \biggl( 
                    3 \Blue{\cf} - \Blue{\ca} 
                 \biggr)
         \biggr] \Red{\ln^2(N)}
       - \biggl[
                 \biggl( 
                    3 \Blue{\cf} - \Blue{\ca} 
                 \biggr)
                 \ln\frac{M^2_H}{\mu^2}
         \biggr] \Red{\ln(N)} \\ \nonumber
     & + \biggl\{ 
            \frac{\Blue{\ca}}{4} 
            \biggl( 3 \Blue{\cf} - \Blue{\ca} \biggr)\z2
          + \biggl[
              \frac{ \Blue{\nf \tr} }{4} 
              \biggl( 2 - \Blue{\cf} \biggr)
            + \frac{\Blue{\ca}}{32} 
              \biggl( 11 \Blue{\cf} - 7 \Blue{\ca} \biggr)
            \biggr] \ln\frac{\mu^2}{m_{\textrm{top}}^2} \\ \nonumber
     &    - \biggl[
            \frac{\Blue{\nf \tr}}{24} 
            \biggl( \Blue{\ca} + 3 \Blue{\cf} \biggr)
          + \frac{5}{48} \Blue{\caca}
            \biggr] \ln\frac{M^2_H}{\mu^2}
          + \frac{\Blue{\nf \tr}}{32} \biggl(
            3 \Blue{\cf} - \frac{17}{3} \Blue{\ca} - 8
            \biggr) \\ \nonumber
     &    + \frac{\Blue{\ca}}{96} \biggl(
            145 \Blue{\cf} + \frac{5}{2} \Blue{\tr} 
                           - \frac{223}{12} \Blue{\ca}
            \biggl)
          + \frac{\Blue{\cf}}{8} 
                           \biggl( \frac{1}{3} \Blue{\tr} 
                                 - \frac{9}{2} \Blue{\cf} \biggr)
         \biggr\} \\
     & + \frac{\Blue{\ca}}{4} 
         \biggl( 3 \Blue{\cf} - \Blue{\ca} \biggr)\z2 .
\end{align}

Finally, as we have the NLO corrections for each of the processes, we
can compute the $B^{(2)}_{g}$ coefficients for each of them. The scalar
case matches its value in the literature\cite{Catani:2003zt} and the
pseudoscalar result is new. They are

\begin{align}
B^{(2),H}   &= \Blue{\caca} \biggl( \frac{23}{24} + \frac{11}{3} \z2 -
                                    \frac{3}{2} \z3
                            \biggr)
             + \Blue{\nf \tr \cf}
             - \Blue{\nf \tr \ca} \biggl( \frac{1}{6} + \frac{4}{3} \z2
                                  \biggr)
             - \frac{11}{18} \Blue{\cf \ca}, \\
B^{(2),A^0} &= \Blue{\caca} \biggl( \frac{1}{2} + \frac{11}{3} \z2 - 
                                    \frac{3}{2} \z3
                            \biggr)
             + \frac{1}{2} \Blue{\nf \tr \cf}
             - \Blue{\nf \tr \ca} \frac{4}{3} \z2.
\end{align}

\section{Results and Conclusions}
\label{results}

\begin{figure}
  \begin{center}
    \begin{tabular}{c}
      \includegraphics{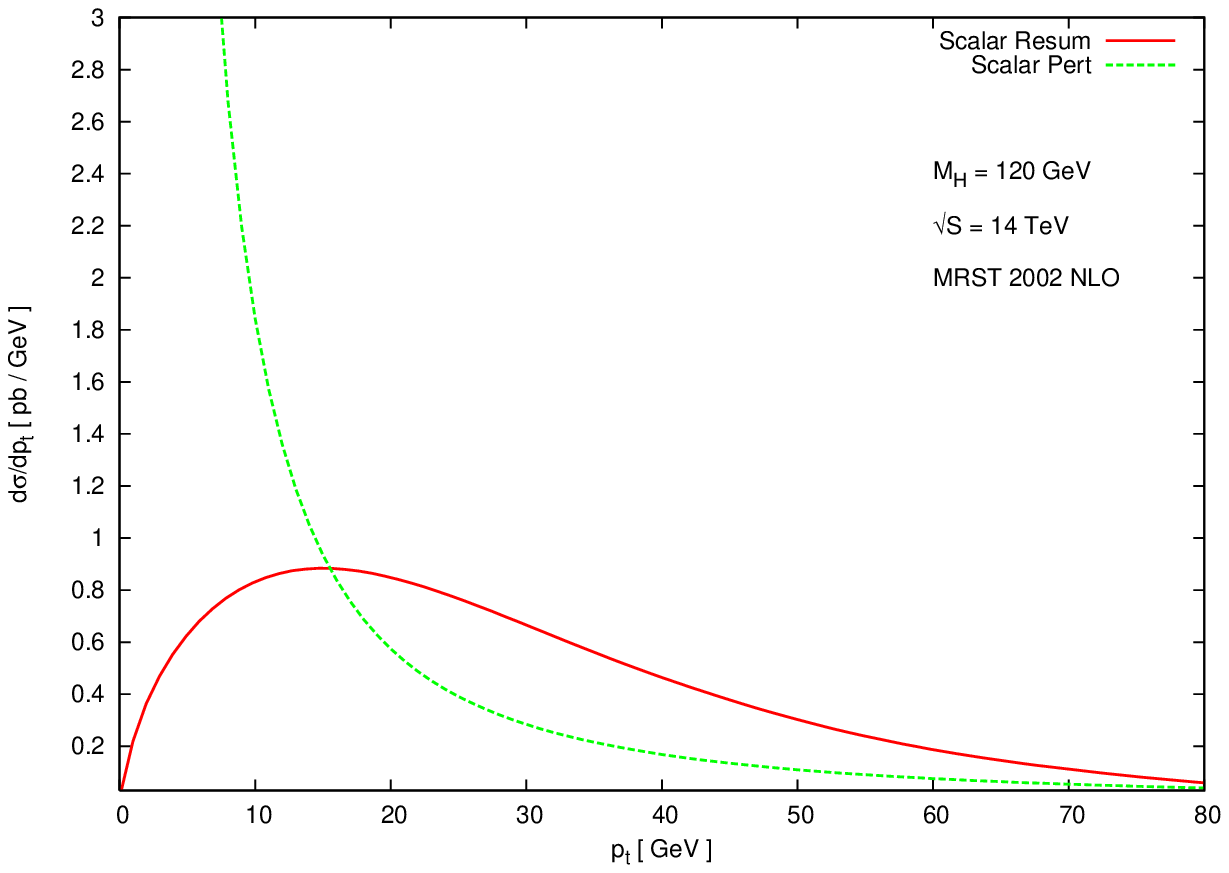} \\
      \includegraphics{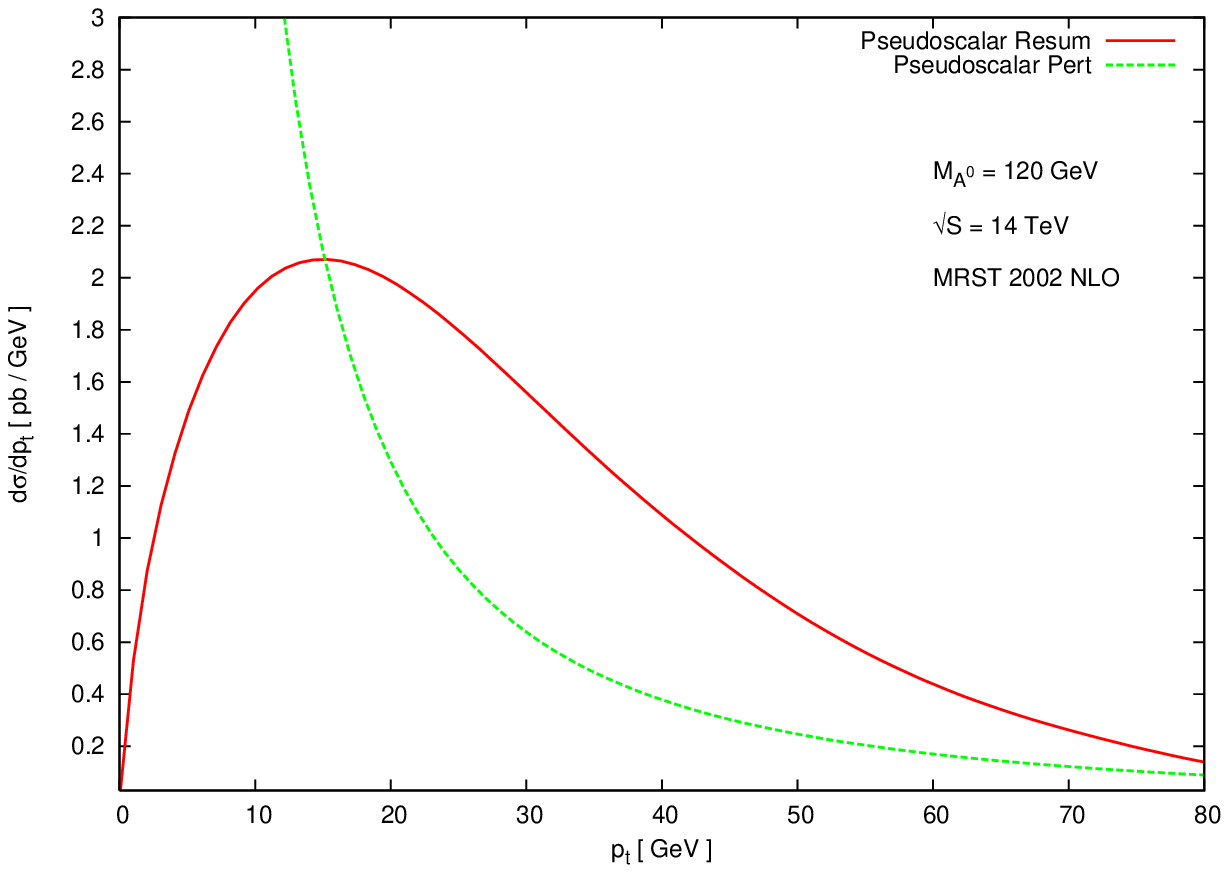}
    \end{tabular}
  \end{center}

\caption{The transverse momentum spectrum for the scalar and
pseudoscalar Higgs boson at the LHC for $|y| \leq 2.5$. The $p_t$
distribution peaks at approximately $15$~GeV. The resummed curve is the
NLL resummation, and the perturbative curve is the NLO fixed order
calculation. The NLO fixed order calculation diverges in the negative
direction at small $p_t$. This piece of the differential cross-section
is not shown for clarity. These two curves cross at approximately
$p_t=100$~GeV/c and stay very close thereafter.}

\label{lhc}
\end{figure}

\begin{figure}
  \begin{center}
    \begin{tabular}{c}
      \includegraphics{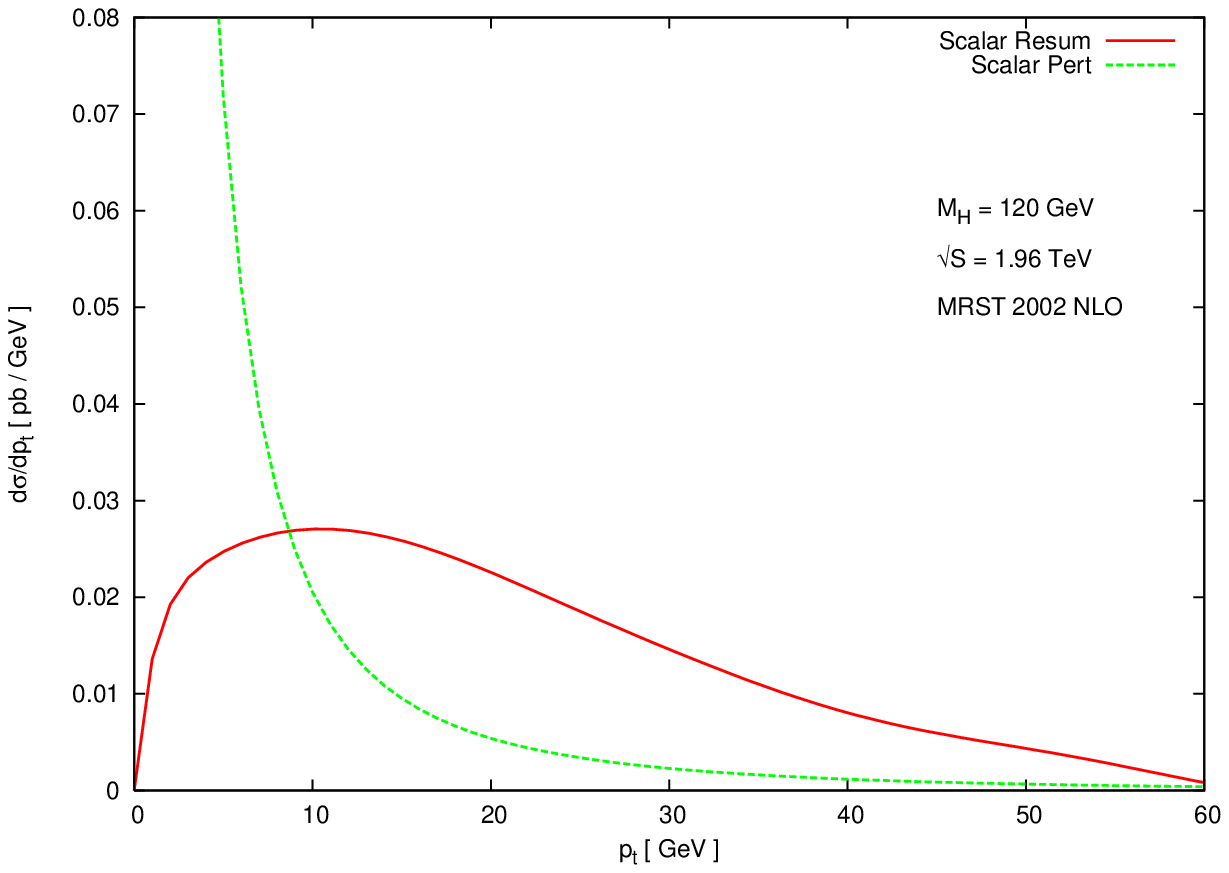} \\
      \includegraphics{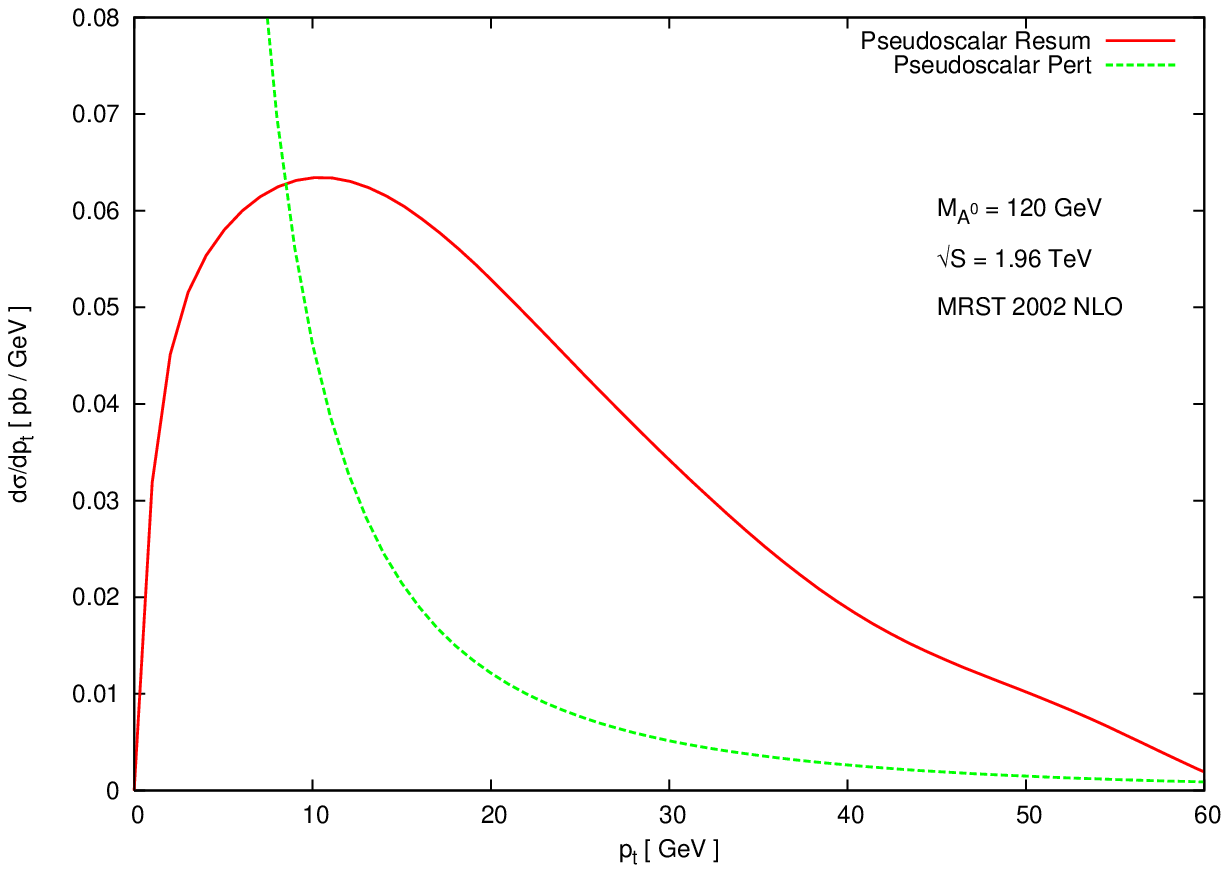}
    \end{tabular}
  \end{center}

\caption{The transverse momentum spectrum for the scalar and
pseudoscalar Higgs boson at the Tevatron for $|y| \leq 2.5$. The $p_t$
distribution peaks at approximately $10$~GeV. The resummed curve is the
NLL resummation, and the perturbative curve is the NLO fixed order
calculation. The NLO fixed order calculation diverges in the negative
direction at small $p_t$. This piece of the differential cross-section
is not shown for clarity. These two curves cross at approximately
$p_t=80$~GeV/c and stay very close thereafter.}

\label{tevatron}
\end{figure}

\begin{figure}
  \begin{center}
      \includegraphics{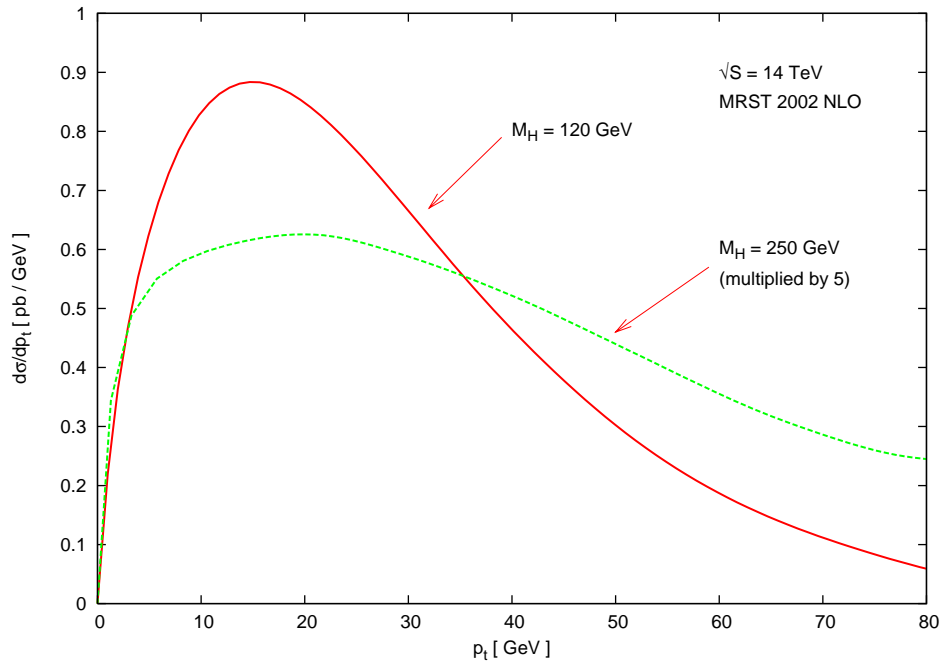}
  \end{center}

\caption{The effects of increasing Higgs mass on the transverse momentum
spectrum at the LHC for $|y| \leq 2.5$. The $M_H = 250$~GeV/c$^2$ curve
peaks at approximately $23$~GeV/c. We can clearly see that the resummed
curve peaks at higher $p_t$ with increasing Higgs mass and that the
width of the resummed distribution becomes wider with increasing Higgs
mass.}

\label{heavy}
\end{figure}

\begin{figure}
  \begin{center}
    \begin{tabular}{c}
      \includegraphics{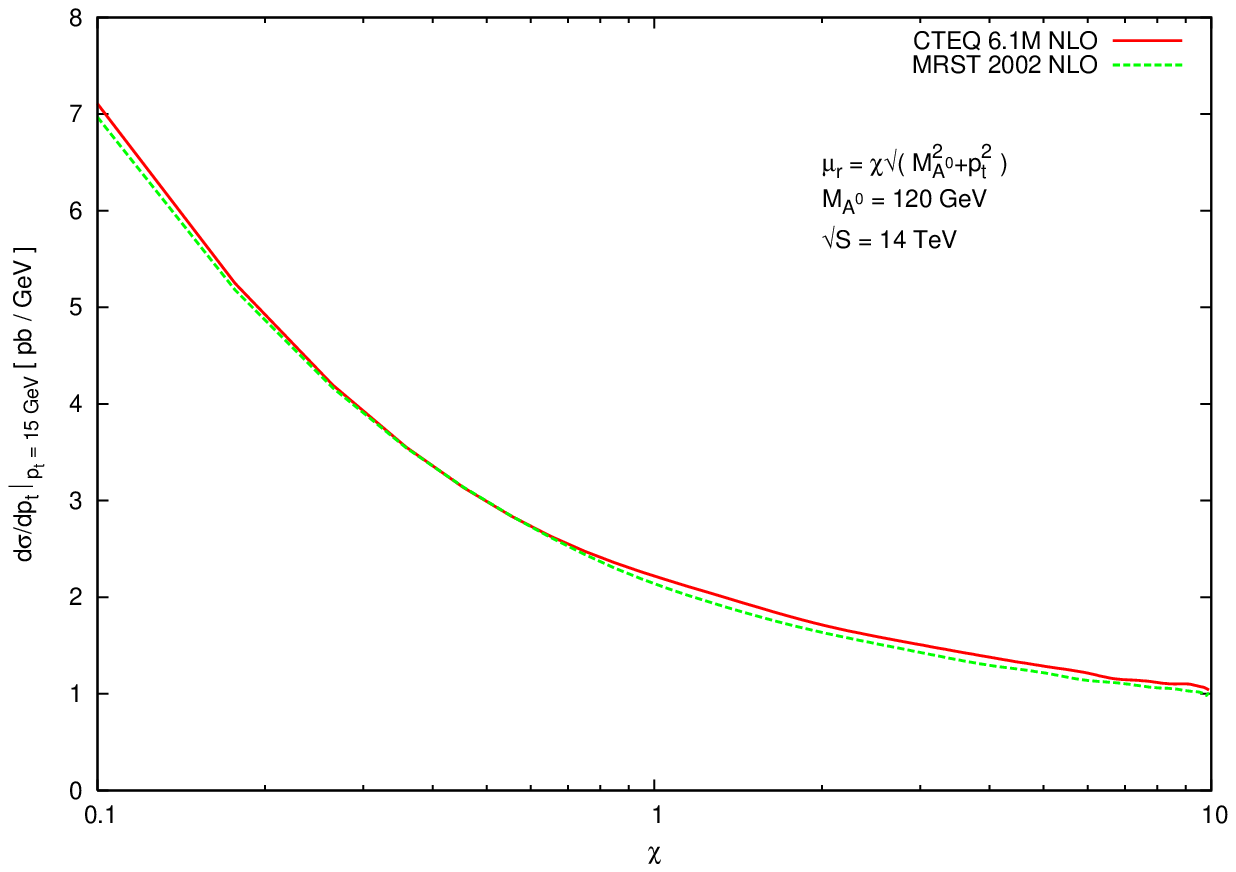} \\
      \includegraphics{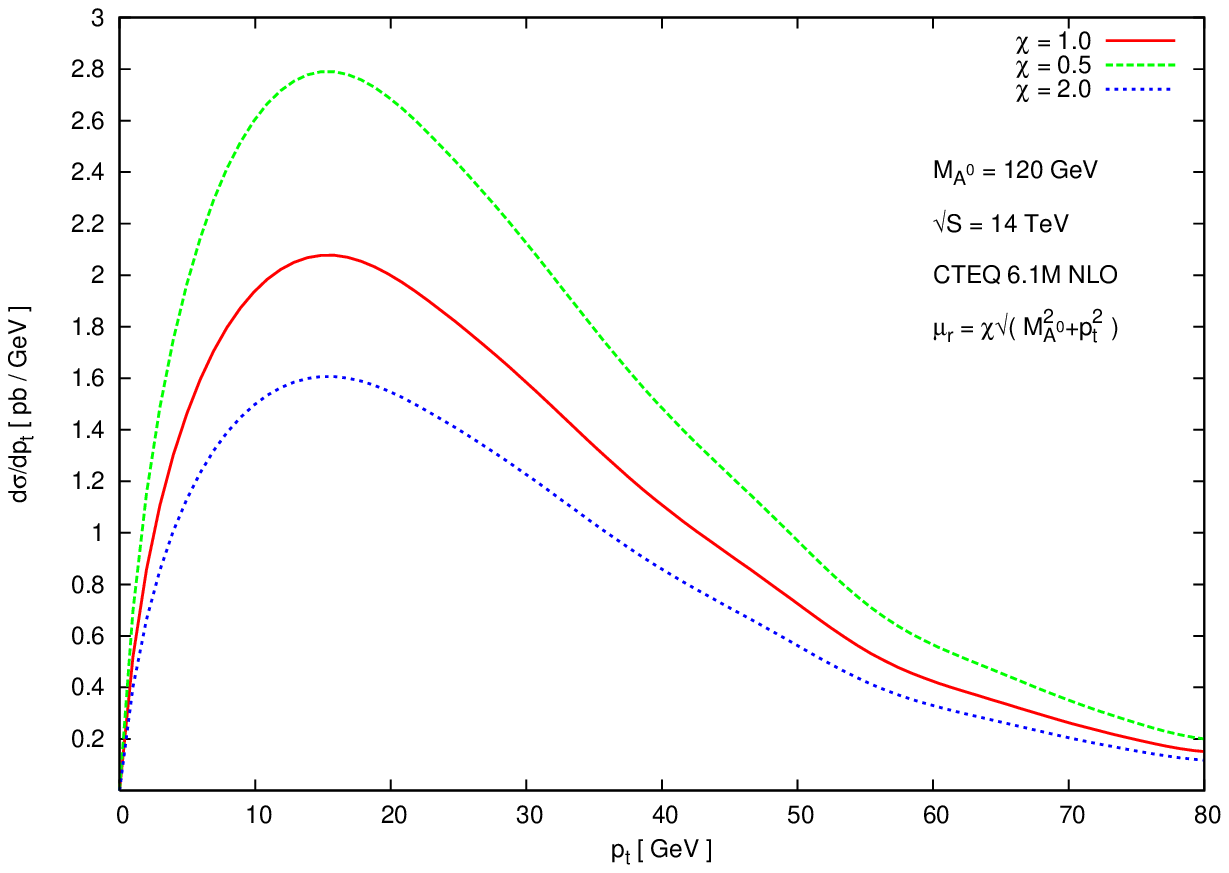}
    \end{tabular}
  \end{center}

\caption{This is the scale uncertainty in the transverse momentum
spectrum for the pseudoscalar Higgs boson. The upper figure shows the
variation on the differential cross-section at its peak near $15$~GeV/c
over a scale variation of an order of magnitude. The upper figure shows
both the MRST and CTEQ parton distribution function. The lower figure
shows the variation over the whole spectrum when the scale is varied by
a factor of two. It is easy to see that the largest scale uncertainty is
at the peak value.
}

\label{scale}
\end{figure}

\begin{figure}
  \begin{center}
    \begin{tabular}{c}
      \includegraphics{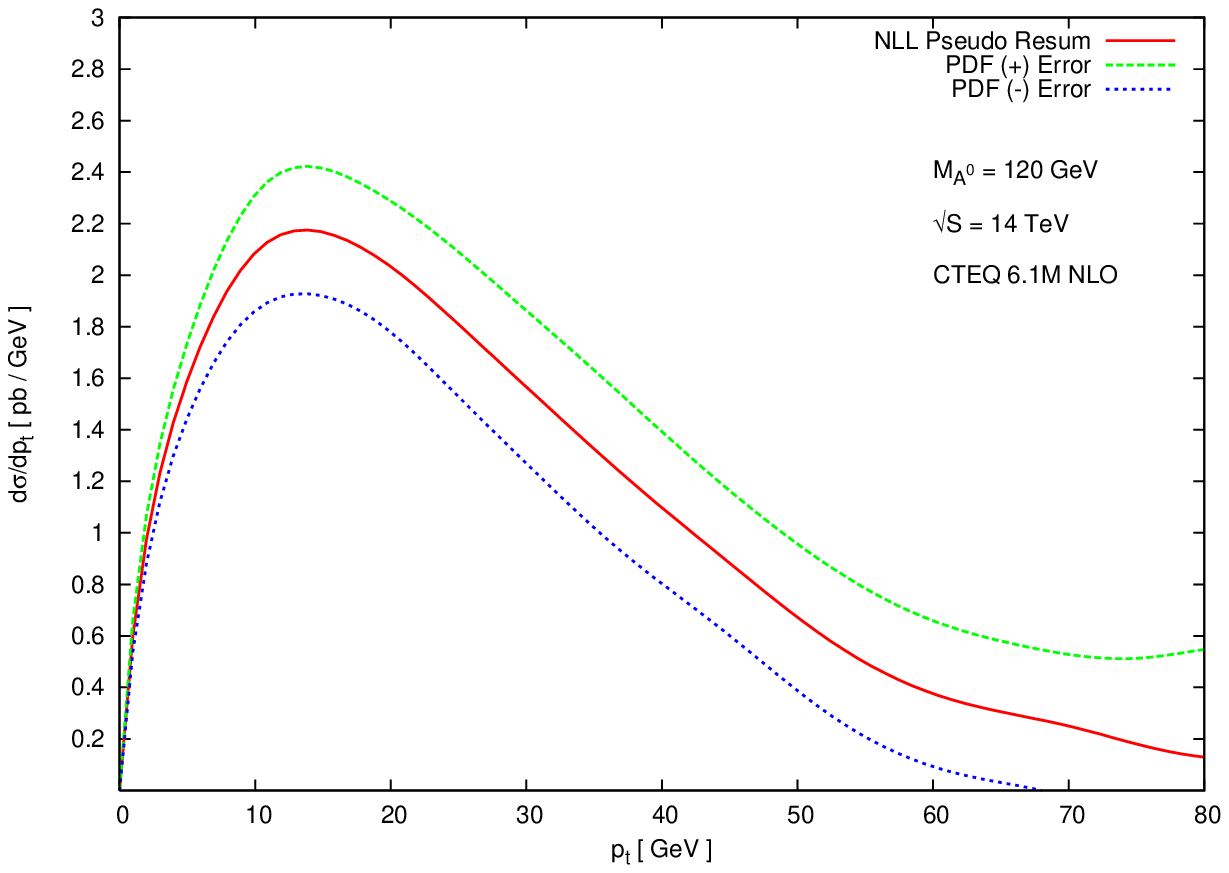}
    \end{tabular}
  \end{center}

\caption{The uncertainty due to the parton distribution functions for
the CTEQ 6.1M parton distribution functions. The resummation is done for
the $gg$ initial state only and therefore has the largest PDF
uncertainty. A $q\bar{q}$ initial state would have a smaller PDF
uncertainty.
}

\label{pdf}
\end{figure}

Although we have shown the explicit differences in the scalar and
pseudoscalar functions, they are numerically quite small. Also, as the
differences in the corrections become greater, they are suppressed more
in $\as$, leaving the predominant difference in the resummed
cross-section is the same factor of $9/4$ that appears in the LO
cross-section. We are also interested in where the resummed $p_t$
distribution peaks at the LHC and Tevatron, so we integrated the
differential cross-sections numerically for a rapidity $|y| \leq 2.5$.

We implemented the MRST2002 NLO updated parton distribution
functions\cite{Martin:2002aw, Martin:2003sk} and the MRST2001 LO parton
distribution functions\cite{Martin:2002dr} in our analysis as well as
the CTEQ 6.1M NLO parton distribution functions\cite{Pumplin:2002vw,
Stump:2003yu}. We have taken the renormalization and factorization
scales to be identical and set equal to $\mu^2 = M_\Phi^2 + p_t^2$. It
should be noted that this choice of scale suppresses the width of the
resummation peak as $p_t$ grows due to the running of the coupling
constant becoming smaller as the scale increases, but the effect is not
a significant one. The LO cross-section ``normalization factor'' in the
resummation formalism also shows $p_t$ dependence for the same reason
when this scale is used.

Our numerical results for the LHC were created with $\sqrt{S}=14$~TeV
and the Tevatron with $\sqrt{S}=1.96$~TeV and a Higgs mass of
$120$~GeV/c$^2$. We used an NLO one-loop $\as(M_Z)=0.1197$ consistent
with the MRST2002 NLO updated parton distribution functions and
$\as(M_Z)=0.118$ for the CTEQ 6.1M NLO parton distribution functions. 
The $p_t$ distributions for the scalar and pseudoscalar Higgs bosons at
the LHC are shown in Figure~\ref{lhc} and Figure~\ref{tevatron} shows
the same figures for the Tevatron. It would appear that in both cases
the factor of $9/4$ difference in the LO cross-sections is the dominant
difference in the small $p_t$ region. The perturbative curves are from
the same computer code that generated the differential cross-sections in
Ref.~\cite{Ravindran:2003um}.

The average transverse momentum and transverse momentum squared at the
LHC for a $120$~GeV/c$^2$ scalar and pseudoscalar Higgs boson with $p_t
= 0-80$~GeV/c are $<p_t> \simeq 27.5$~GeV/c and $<p_t^2>^{1/2} \simeq
32.7$~GeV/c with a peak value at $15$~GeV/c. At the Tevatron the average
transverse momentum and transverse momentum squared for the scalar and
pseudoscalar Higgs boson with $p_t = 0-60$~GeV/c are $<p_t> \simeq
20$~GeV/c and $<p_t^2>^{1/2} \simeq 24$~GeV/c with a peak value at
$10$~GeV/c. These values should be used with caution as they depend on
the PDFs used in the analysis.

We are interested in what happens when a much heavier Higgs boson is
considered. This would be the case if one were interested in a heavy SM
Higgs, the heavy scalar $H^0$ in the MSSM, or a heavy pseudoscalar
Higgs. We also ran our code for a Higgs with a mass of $250$~GeV/c$^2$. 
We found a few interesting trends. The peak in the differential
distribution moved to a higher $p_t$ as expected\cite{Berger:2002ut} and
the width of the peak became much broader. The width of the peak is
interesting because it is telling us something about the decay width for
the Higgs. The scale of the cross-section also dropped considerably as
one would expect for a heavier final state particle. As the mass of the
Higgs became very heavy, it became hard to distinguish a discernable
peak in the distribution as it became very wide. As we can see in
Figure~\ref{heavy}, this is a pronounced effect. The resummed curve
becomes so broad that it does not cross the fixed order differential
cross-section until very high transverse momentum.

There is a great deal of interest in understanding the uncertainties
associated with the differential cross-section. Although it is common to
look at the scale dependence of a total cross-section for scalar Higgs
production\cite{Djouadi:2003jg}, we would like to see how our results
are effected by changes in the scale factor $\mu$ and the uncertainty in
the parton distribution functions for the differential cross-section.

In Figure~\ref{scale}, the upper graph shows the scale dependence of the
peak of the distribution when the scale factor is varied by a factor of
ten. The lower graph in the same figure shows how the entire
distribution changes when the scale is changed by a factor of two. From
this lower graph is it easy to see that the peak of the distribution has
the most sensitivity to the scale parameter. We define a prefactor to
our renormalization scale to allow it to be varied with ease. We define
$\mu^2 = \chi^2(M_\Phi^2 + p_t^2)$. When the scale factor is changed by
a factor of ten lower ($\chi = 0.1$) the peak increased by a factor of
approximately $3.1$ and when the scale factor is increased by a factor
of ten higher ($\chi = 10$) the peak of the distribution is lowered by
factor of approximately $0.46$. When the scale is only varied over a
more reasonable factor of two, then the peak moves by approximately
$25\%$. The overall scale dependence is very close to $\as^2(\mu)$
running as expected from the $\sigma_0^{\text{LO}}$ prefactor in the
resummation formalism. Overall, we can see that the shape of the
distribution is not effected greatly by the change in the scale
parameter, only its magnitude is changed significantly.

It is well known that the CTEQ gluon distribution is higher at small $x$
than the MRST sets which can be seen in the upper graph in
Figure~\ref{scale}, but the effect is quite small. Otherwise, the two
distributions are very similar and can be considered interchangable in
this analysis.

In Figure~\ref{pdf}, the uncertaity due to the parton distribution
functions is shown. At the peak of the distribution, we see an
uncertainty of approximately $10\%$. Considering only scale variations
of a factor of two would lead us to believe that there is still
approximately a $35\%$ uncertainty in the differential cross-section at
its peak. The uncertainty would be slightly lower at other values of the
transverse momentum due to the scale $\mu$ and larger at higher values
of the transverse momentum due to the PDF uncertainty.

In this paper, we have calculated the resummation coefficients for
pseudoscalar Higgs boson production for both the total cross-section,
presenting the $B^{(2)}_g$, $C^{(1)}_{gg}$, and $C^{(2)}_{gg}$
coefficients, and the differential cross-section, presenting the
$\bar{C}^{(1)}_{gg}$ coefficient. We have also shown the effects of
increasing the mass of the Higgs boson on the resummed differential
cross-section and performed an analysis of the uncertainties associated
with the renormalization scale and the parton distribution functions.

\begin{acknowledgments}

I would like to thank J.~Smith, S.~Dawson, J.~Vermaseren, W.~Vogelsang,
G.~Sterman, N.~Christensen, and A.~Field-Pollatou for all their help and
comments during the several stages of this paper. The author is
supported in part by the National Science Foundation grant PHY-0098527.

\end{acknowledgments}

\appendix 
\section{Harmonic Polynomials}

Finding the Mellin moments of the fixed order total cross-section
corrections has been made considerably simpler with the \textsc{harmpol}
package in \textsc{form}\cite{Vermaseren:2000nd}. In order to use this
powerful package, it is necessary to express the polylogarithmic
expressions in terms of harmonic polylogarithms\cite{Vermaseren:1998uu,
Remiddi:1999ew}.

Harmonic polylogarithms are defined recursively in three classes for
each weight. To make this clear, let us define three functions

\begin{equation}
f(-1;x)=\frac{1}{1+x}, \quad
f(0;x)=\frac{1}{x}, \quad
f(1;x)=\frac{1}{1-x},
\end{equation}
so we can define the weight $w=1$ harmonic polylogarithms as
\begin{equation}
\h(a;x)=\int_0^x dx' \, f(a;x').
\end{equation}
Thus the first three harmonic polylogarithms can be written explicitly 
as
\begin{equation}
\h(-1;x) = \ln(1+x), \quad
\h(0;x) = \ln(x), \quad
\h(1;x) = -\ln(1-x).
\end{equation}
For higher weight harmonic polylogarithms, we need to generalize the 
notation. The $w$-dimensional vector $\vec{m}_w$ should be broken into 
the first index and the rest of the vector as 
$\vec{m}_w=(a,\vec{m}_{w-1})$. This gives us a general expression 
for the rest of the harmonic polylogarithms recursively,
\begin{equation}
\h(\vec{0}_w;x)=\frac{1}{w!}\ln^w \! x, \quad
\h(\vec{m}_w;x)=\int_0^x dx' \, f(a;x') \, 
\h(\vec{m}_{w-1};x').
\end{equation}

Although it is easy to find the harmonic polylogarithmic expression for
the logarithms and dilogarithms, some further work is needed for the
dilogarithms with quadratic arguments and the trilogarithms that appear
in the NNLO corrections. The dilogarithms can be simplified in a very
straightforward way using well known relationships. To list them briefly
the most useful expressions are

\begin{align}
-\li2(1-x^2) &= \, 2[\li2(x) + \li2(-x)
                   + \ln(x) \ln(1-x^2)] - \z2 \\
-\li2(1-x)   &= \, \li2(x) + \ln(x) \ln(1-x) - \z2 \\
 \li2(x)     &= \, \h(0,1;x)  = \h_{2}(x)\\
-\li2(-x)    &= \, \h(0,-1;x) = \h_{-2}(x)
\end{align}

Fewer relationships exist for the trilogarithms. It proved to be very
challenging to remove three of the trilogarithmic expressions
simultaneously from the NNLO corrections. The following expressions were
derived from the polylogarithm literature\cite{Lewin:1958, Lewin:1981}
and are presented here for future reference (using the notation for the
Harmonic polylogarithms of Ref.~\cite{Remiddi:1999ew}). That allows one
to express the NNLO corrections completely in terms of Harmonic
polylogarithms,
\begin{align} \nonumber
\li3 \biggl( \frac{+(1-x)}{1+x} \biggr) -
      \li3\biggl( \frac{-(1-x)}{1+x} \biggr) =& \,
    2 \li3(1-x) + 2 \li3\biggl( \frac{1}{1+x} \biggr)
  - \frac{1}{2} \li3(1-x^2), \\
 &- \frac{7}{4}\z3 + \z2 \ln(1+x) - \frac{1}{3} \ln^3(1+x), \\ \nonumber
\li3 \biggl( \frac{x}{1+x} \biggr) =& \, 
    \frac{1}{6} \biggl[ \ln^3 \biggl( \frac{1+x}{x} \biggr) 
  + \ln^3(1+x) \biggr] - \biggl( \frac{\z2}{2} + \frac{\ln^2(x)}{4} 
    \biggr) \biggr[ 2\ln(1+x)-\ln(x) \biggl] \\ 
  & \quad - \frac{1}{2} \biggl[ \li3 \biggl( -\frac{1}{x} \biggr) 
  + \li3(-x) \biggr] - \li3 \biggl( \frac{1}{1+x} \biggr) 
  + \z3, \\ \nonumber
\li3 \biggl( \frac{1}{1+x} \biggr) =& \, 
    \frac{1}{2} \ln^2(1+x)\ln(x) - \ln(1+x) \biggl[ \li2(-x) 
  + \ln(x)\ln(1+x) - \z2 \biggr] + \z2 \\
 &- \h_{-2,-1}(x) + \z2 \ln(1+x) + \frac{1}{6} 
    \biggl[ \ln^3(1+x) - 18\z2\ln(1+x) \biggr], \\ \nonumber
\li3 \biggl( \frac{-(1-x)}{x} \biggr) =& \,
  -\li3(1-x) -\li3(x) + \z3 + \z2\ln(1-x) - \frac{1}{2}\ln(x)\ln^2(1-x)
  + \frac{1}{6} \ln^3(1-x) \\ 
 &- \z2 \ln\biggl( \frac{1-x}{x} \biggr)
  - \frac{1}{6} \ln^3 \biggl( \frac{1-x}{x} \biggr), \\ \nonumber
\li3 \biggl( \frac{-(1-x^2)}{x^2} \biggr) =& \,
  - \li3(1-x^2) - 4 \biggl[ \li3(x) + \li3(-x) \biggr] + \z3
  + \z2  \ln(1-x^2) \\ 
 &- \ln(x) \ln^2(1-x^2) + \frac{1}{6} \ln^3(1-x^2)
  - \z2 \ln \biggl( \frac{1-x^2}{x^2} \biggr)
  - \frac{1}{6} \ln^3 \biggl( \frac{1-x^2}{x^2} \biggr), \\
\li3(1-x^2) =& \,  \ln(x)\ln^2(1-x^2) + \li2(1-x^2) \ln(1-x^2) 
  + \z3 -2 \biggl[ \h_{2,1}(x)-\h_{-2,1}(x) \biggr], \\
\li3(1-x) =& \, \frac{1}{2} \ln(x) \ln^2(1-x)
  + \ln(1-x) \li2(1-x) + \z3 - \h_{2,1}(x), \\
\li3 \biggl( -\frac{1}{x} \biggr) =& \,
    \li3(-x) + \z2 \ln(x) + \frac{1}{6} \ln^3(x), \\
 \li3(x)  =& \, \h_{3}(x), \\
-\li3(-x) =& \, \h_{-3}(x), \\
-S_{1,2}(1-x) =& \, \li3(x) + \ln(x) \li2(x)
                  + \frac{1}{2} \ln(1-x) \ln^2(x) + \z3, \\
-S_{1,2}(-x)  =& \, \h_{-2,-1}(x).
\end{align}

It was also necessary to linearize all the arguments of the natural logs
and to partial fraction the inverse powers of $1-x^2$. With these above
expressions, it was possible to use \textsc{harmpol} to find the Mellin
moments of the correction factors. Although most of these expressions
were verified numerically, it should be emphasized that these
expressions were derived so that they would be valid at $x \leq 1$,
which is where they would be evaluated on threshold. In some regions
these expressions would pick up imaginary pieces, but since we are
interested in corrections to a partonic cross-section our expression
must stay real.

Using these expressions, the NNLO corrections in
Refs.~\cite{Harlander:2001is, Harlander:2002vv, Ravindran:2003um} can be
reduced to Harmonic polylogarithms so their moments can be easily found.

\end{document}